\documentclass[aps,prb,twocolumn,showpacs,groupedaddress]{revtex4}
\usepackage{amssymb}
\usepackage{bm}
\usepackage{amsmath}
\usepackage{graphicx}
\usepackage{hyperref}

\begin{document}

\title{Phenomenological $Z_2$ lattice gauge theory of the spin-liquid state of the kagome Heisenberg antiferromagnet}

\author{Yuan Wan}
\author{Oleg Tchernyshyov}
\affiliation{Department of Physics and Astronomy, The Johns Hopkins University, Baltimore, Maryland 21218}

\date{\today}


\begin{abstract}
We construct a phenomenological $Z_2$ lattice gauge theory to describe the spin-liquid state of the $S=1/2$ kagome Heisenberg antiferromagnet in the sector with zero total spin. The model is a natural generalization of the Misguich-Serban-Pasquier Hamiltonian with added interactions that are strongly constrained by lattice symmetries. We are able to reproduce qualitatively many of the characteristic features observed in recent numerical studies. We also make connections to the previous works along similar lines.
\end{abstract}

\maketitle

\section{Introduction}

Quantum spin liquids (QSLs) are symmetry-preserving quantum phases of matter hosting fractionalized excitations. \cite{Balents} They not only possess unusual physical properties, which makes them interesting from the perspective of fundamental physics, but may also provide new opportunities for quantum information technology. \cite{Kitaev} The $S=1/2$ kagome Heisenberg antiferromagnet (KHAF) has emerged as a promising candidate system from a decades-long search for QSL in theoretical models and real materials. Kagome is a lattice of corner-sharing triangles. Spins, residing in the vertices of kagome and interacting antiferromagnetically, are frustrated by the inability to minimize all pairwise interactions. The strong frustration suppresses spontaneous symmetry breaking and thus opens the possibility of a spin-liquid state. \cite{Ramirez} The $S=1/2$ KHAF is realized in several materials, most notably in herbertsmithite ZnCu$_3$(OD)$_6$Cl$_2$, which shows signs of a QSL. \cite{Han}

The Hamiltonian of the $S=1/2$ KHAF is 
\begin{align}
H=\sum_{\langle{}ij\rangle}\mathbf{S}_i\cdot\mathbf{S}_j,\label{khaf}
\end{align}
where $\mathbf{S}_i$ are $S=1/2$ spin operators defined on vertices of kagome, and the summation is over nearest-neighbors (Fig.~\ref{dimer_covering}a). Despite its simplicity, determining the ground state of Eq.~(\ref{khaf}) has proved to be a formidable task. Several competing proposals for the nature of its ground state have been made over the past two decades. \cite{Sachdev, Hastings, Singh, Ran, Evenbly, Yan, Depenbrock, Messio, Iqbal, Capponi}

Recent numerical studies based on the density-matrix renormalization group (DMRG) method cast new light on this problem. They provide compelling evidence for a QSL. In the ground state, spins have very short-ranged correlations, gapped $S=1$ excitations, and no signs of long-range magnetic order. \cite{Yan, Depenbrock} The ground-state wave function bears topological entanglement entropy with total quantum dimension $D=2$, hinting at a spin liquid of the $Z_2$ type. \cite{Depenbrock, Jiang}  Additionally, the numerical studies have revealed a wealth of new features that remain to be understood, including a strong valence-bond resonance around diamond-shaped loops \cite{Yan} and peculiar valence-bond correlations near lattice defects \cite{White}. The ground state of Eq.~\ref{khaf} with cylindrical geometry shows a strong dependence on the circumference and the chirality of the cylinder. \cite{Yan} Localized $S=1/2$ spins are found on the open edges of certain types of cylinders. \cite{Jiang2} These numerical findings call for a theoretical assessment.

On the theory front, much effort has been made to construct a low-energy effective description of the $Z_2$ spin liquid phase of the $S=1/2$ KHAF. A $Z_2$ spin-liquid theory is commonly obtained by mapping a quantum dimer model (QDM), itself a low-energy description of the $S=1/2$ Heisenberg model, to a $Z_2$ gauge theory. This method was initiated by \textcite{Moessner2} and has been successfully applied to the $Z_2$ spin-liquid phase of the triangular antiferromagnet\cite{Moessner1} and to the $J_1$--$J_2$ Heisenberg model on a square lattice.\cite{J1J2} On kagome, an exactly solvable QDM with a $Z_2$ spin-liquid ground state was constructed by \textcite{Misguich}. Its $Z_2$ flux (vison) excitations are gapped and completely localized. Dynamical visons were introduced by \textcite{Nikolic} and by \textcite{Huh}, and various $Z_2$ spin-liquid phases and proximate valence-bond crystals were classified on formal grounds. Yet, how the aforementioned numerical results fit into a concrete $Z_2$ spin liquid theory remains an open question.

In this paper we provide a phenomenology of the $Z_2$ spin-liquid phase of the $S=1/2$ KHAF, which bridges the theory and the numerics. We demonstrate that the kagome QDM is naturally described in the language of a $Z_2$ gauge theory. We construct a  $Z_2$ lattice gauge theory and show that the model indeed reproduces qualitatively various features of the ground state of the $S=1/2$ KHAF revealed by the DMRG studies.

The paper is organized as follows. In Section~\ref{Model} we describe the construction of the phenomenological model. In Section~\ref{Method} we describe techniques for solving the phenomenological model. In Section~\ref{Results} we solve the model in various settings and compare the solutions with the numerical results. In Section~\ref{Discussion} we discuss the connection between the current work with previous works.

\section{Construction of the model}\label{Model}

In this section we construct a phenomenological model describing the $Z_2$ spin liquid phase of the $S=1/2$ kagome antiferromagnet. Our starting point is a QDM on kagome that is thought to represent low-energy states of the $S=1/2$ KHAF in the sector with zero total spin. We first show that the Hilbert space of kagome QDM is identical to that of a $Z_2$ gauge theory on a dual lattice, a honeycomb whose sites are centers of kagome triangles. We recast the QDM Hamiltonian of Misguich, Serban, and Pasquier (MSP), known to have a $Z_2$ spin-liquid ground state, as a $Z_2$ gauge theory. We then perturb the MSP Hamiltonian by adding dimer interactions consistent with the lattice symmetry. We keep only the leading terms beyond the MSP model. The resulting $Z_2$ gauge theory is mapped onto a dual Ising model in a transverse field, which is used to compute the density and correlations of quantum dimers. 

\subsection{From quantum dimer model to $Z_2$ gauge theory}\label{qdm2z2}

\begin{figure}
\includegraphics[width=\columnwidth]{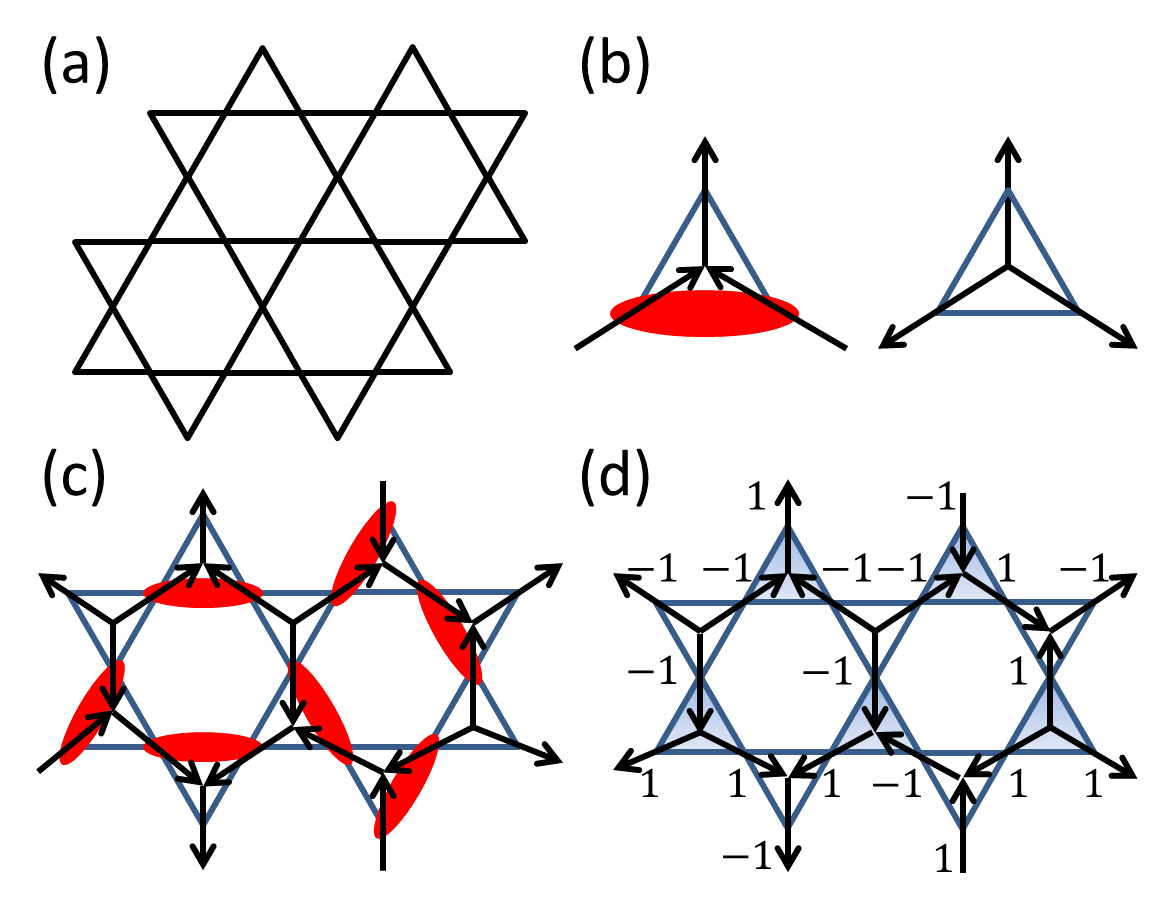}
\caption{(a) The kagome Heisenberg antiferromagnet. $S=1/2$ spins are located on the lattice sites, and the antiferromagnetic Heisenberg interactions are between nearest-neighboring spins. (b) The rules of the arrow representation of the dimer coverings. (c) An example of the arrow representation. (d) Mapping the arrow patterns to Ising variables. Shaded (unshaded) triangles are A (B) sites of the honeycomb.}\label{dimer_covering}
\end{figure}

The starting point is a QDM on kagome introduced by Zeng and Elser. \cite{Zeng} A dimer represents two adjacent $S=1/2$ spins in a state with total $S=0$. The QDM Hamiltonian acts on the space of dimer coverings, states in which each site forms a dimer with one nearest neighbor.  The basic assumption of the QDM is that dimer coverings span the low-energy subspace of Eq.\ref{khaf} with total spin $S=0$. \textcite{Elser} mapped dimer coverings on kagome to patterns of arrows connecting centers of adjacent triangles (and thus forming a honeycomb). On a triangle with a dimer, one arrow points out of the triangle and away from the dimer; the other two arrows point into the triangle. A triangle without a dimer has three arrows pointing out. The mapping is illustrated in Fig.~\ref{dimer_covering}b. It is easy to check that the arrows are subject to a constraint: each vertex of the honeycomb lattice has either one outgoing and two incoming arrows, or three outgoing arrows. An example of such a mapping is shown in Fig.~\ref{dimer_covering}c. (This convention, used by \textcite{Misguich}, is in fact opposite to the one used by \textcite{Elser}.)

We introduce Ising variables $\sigma_{ij}^{x}\equiv\sigma^x_{ji}=\pm1$, defined on the links of the honeycomb lattice, to parametrize the states of the arrows. To this end, we partition the honeycomb sites into A and B sublattices. For a given honeycomb link $\langle{}ij\rangle$, $\sigma_{ij}^x=1$ if the arrow points from an A site to a B site, and $-1$ otherwise (Fig.~\ref{dimer_covering}d). The constraint on arrows is thereby translated to a constraint on Ising variables:
\begin{align}
{}Q_{i}\equiv\sigma^x_{i1}\sigma^x_{i2}\sigma^x_{i3}=\left\{
\begin{array}{cc}
1 & i\in{}A\\
-1 & i\in{}B
\end{array}
\right.{}.\label{constraint}
\end{align}
Here $i$ denotes a site on the honeycomb lattice, whereas $\langle{}i1\rangle$,$\langle{}i2\rangle$, and $\langle{}i3\rangle$ are the three links emanating from the site $i$.

We interpret the variables $\sigma^{x}_{ij}$ as the electric flux operators of the $Z_2$ gauge theory on a honeycomb. The constraint Eq.~\ref{constraint} then becomes Gauss's law, and $Q_i$ is the $Z_2$ charge on a honeycomb site $i$.

We have thus established that the Hilbert space of a kagome QDM is identical to that of a $Z_2$ gauge theory on a honeycomb with staggered background charges (\ref{constraint}). As a result, any kagome QDM Hamiltonian can be written as a $Z_2$ gauge theory Hamiltonian.

\subsection{MSP Hamiltonian as $Z_2$ gauge theory Hamiltonian}

\begin{figure}
\centering
\includegraphics[width=\columnwidth]{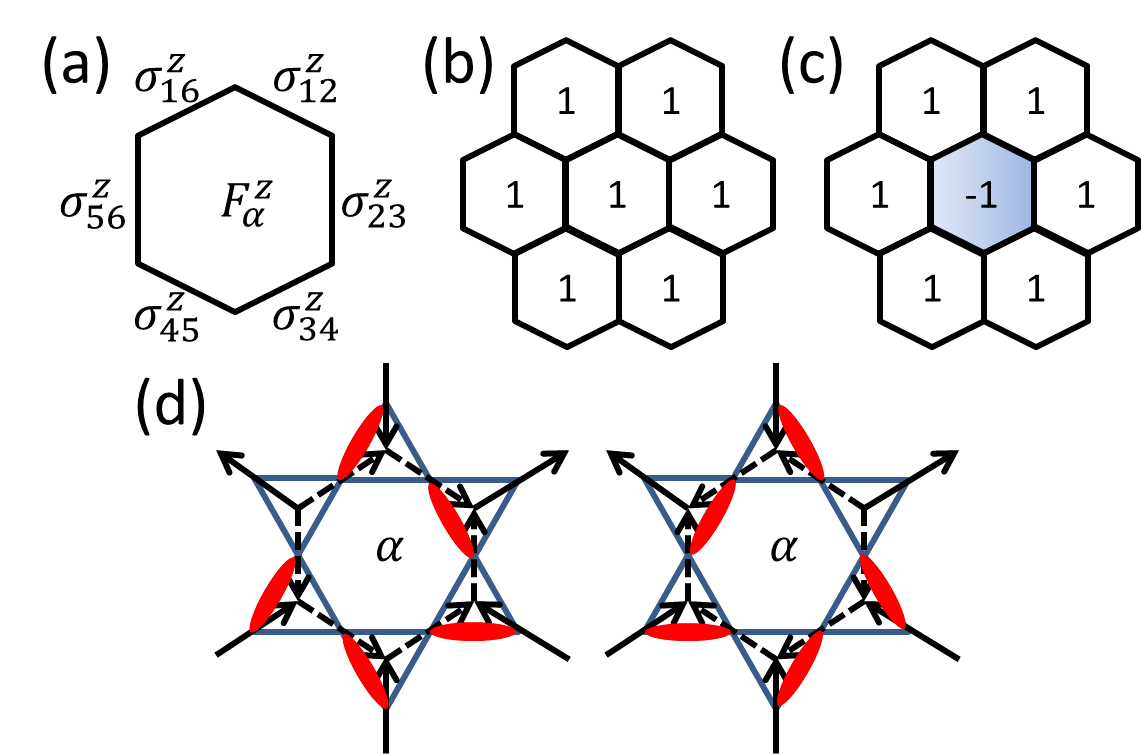}
\caption{(a) The flux $F^z$ is defined as the product of $\sigma^z_{ij}$ around the hexagon. (b) The magnetic flux threading any hexagon is $1$ in the ground state. (c) A vison state contains a $-1$ flux threading through a plaquette (shaded hexagon).(d) The effect of $F^z_\alpha$ on the electric fluxes is to reverse all the arrows around the hexagonal plaquette $\alpha$, or equivalently flipping the dimers in a David Star.}\label{flux_vison}
\end{figure}

The MSP Hamiltonian describes the simplest, exactly solvable QDM on kagome. \cite{Misguich} Its ground state is an equal-amplitude superposition of all possible dimer coverings, the analog of the Kivelson-Rokhsar state \cite{Rokhsar} on kagome. Here we will motivate the MSP Hamiltonian from the gauge theory perspective.

In Section \ref{qdm2z2}, we have introduced the electric flux operator $\sigma^{x}_{ij}$ and related it to the arrow orientation on link $\langle{}ij\rangle$. Here we consider the $\mathbb{Z}_2$ magnetic flux operator associated with a hexagonal plaquette of the honeycomb (Fig.~\ref{flux_vison}a):
\begin{align}
F^z_{\alpha}=\prod_{\langle{}ij\rangle\in\alpha}\sigma^z_{ij}=\sigma^z_{12}\sigma^z_{23}\sigma^z_{34}\sigma^z_{45}\sigma^z_{56}\sigma^z_{61}.
\end{align}
Here $1,2,\dots{}6$ are sites in hexagon $\alpha$, and $\sigma^z_{ij}$ are the standard $Z_2$ link variables. $F^z_\alpha$ commutes with charge operators $[F^z_{\alpha},Q_i]=0$, $\forall\alpha,{}i$. Therefore, the Hilbert space subject to the charge constraint Eq.~\ref{constraint} is invariant under $F^z_{\alpha}$. Electric flux operators $\sigma^{x}_{ij}$ and magnetic flux operators $F^z_{\alpha}$ are the building blocks of a $Z_2$ gauge theory \cite{Kogut}.

Consider a simple $Z_2$ gauge theory Hamiltonian, containing only magnetic terms:
\begin{align}
H=-h\sum_{\alpha}F^z_{\alpha},\label{misguich}
\end{align}
The summation is over all hexagonal plaquettes; $h>0$. $F^z_{\alpha}$ reverses all arrows around the hexagon $\alpha$ regardless the initial orientation. In terms of dimers on kagome, $F^z_{\alpha}$ shifts dimers along a closed path within a David star with amplitude $-h$ (Fig.~\ref{flux_vison}d). which is identical to the operator ``$\sigma^x(h)$'' defined by \textcite{Misguich}. Thus, the Hamiltonian Eq.~\ref{misguich} is nothing but the MSP Hamiltonian formulated in the $Z_2$ gauge theory language.

The MSP Hamiltonian Eq.~\ref{misguich} is solvable. The eigenstates are labeled by the magnetic flux threading through every plaquette. The ground state contains no magnetic flux, and it has been shown that the ground state respects all symmetries of the lattice and exhibits short-ranged correlation, as expected for a gapped spin liquid (Fig.~\ref{flux_vison}b). The excited states are created by inserting $-1$ fluxes, known as visons in literature, and the energy cost for each vison is $2h$ (Fig.\ref{flux_vison}c). Visons are completely localized and dispersionless in the MSP Hamiltonian.

\subsection{Building the phenomenological Hamiltonian}

\begin{figure}
\includegraphics[width=\columnwidth]{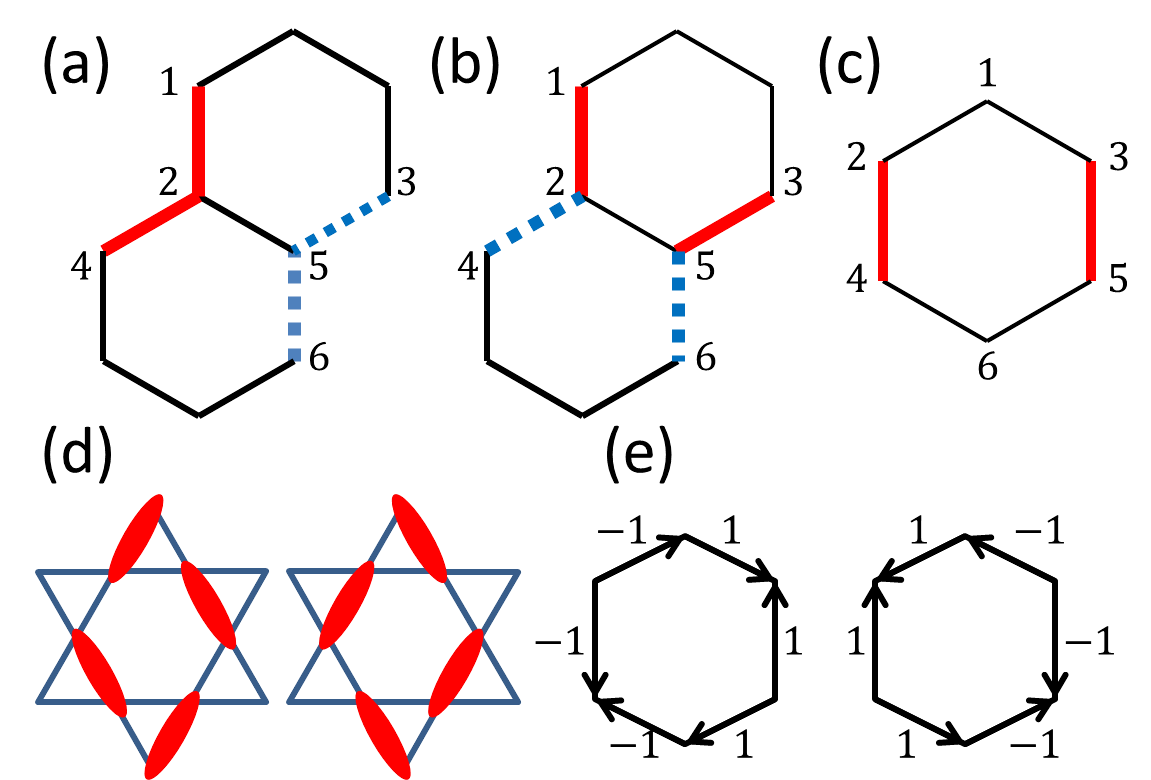}
\caption{(a) The links highlighted in red (thick solid) are a pair of nearest neighbors, and the links highlighted in blue (thick dashed) are another pair. (b) Similar to (a), the links highlighted in the same format are second neighbors. (c) A pair of third neighbor links are highlighted in red (thick solid). (d) The two quasi-degenerate dimer coverings in the ``diamond'' resonance observed by DMRG calculations. (e) The arrow representation for the diamond resonance. }\label{sigma_x_interact}
\end{figure}

The MSP model is unfortunately too simple to account for the rich features observed in DMRG numerics. For instance, the response to a lattice defect is confined to its immediate neighborhood as opposed to the ripple-like features observed in DMRG calculations. \cite{White} Yet, the MSP Hamiltonian provides a natural starting point for the construction of a phenomenological $Z_2$ gauge theory. In what follows, we perturb the MSP model by adding electric flux operators $\sigma^x$ to the Hamiltonian. From the quantum dimer model perspective, this amounts to including dimer density interactions. 

We consider generic interactions up to the second order in the electric fluxes $\sigma_x$: 
\begin{equation}
V = -\sum_{\langle{}ij\rangle}A^{(1)}_{ij}\sigma^{x}_{ij}
	-\sum_{\langle{}ij\rangle,\langle{}kl\rangle}A^{(2)}_{ij;kl}\sigma^x_{ij}\sigma^x_{kl},
\label{generalform}
\end{equation}
The coupling constants $A^{(1)}_{ij}$, $A^{(2)}_{ij;kl}$ must be compatible with the symmetries of the kagome and the honeycomb lattices. 

We first show that the linear terms are not allowed because they are odd under some lattice symmetries. A $\pi$ rotation about a kagome vertex shared by triangles $i$ and $j$ reverses the arrow on honeycomb link $\langle{}ij\rangle$. This alters the sign of the electric flux, $\sigma^x_{ij}\to-\sigma^x_{ij}$, hence $A^{(1)}_{ij} = 0$. Physically, a linear term expresses a preference for some dimer configuration, any of which violates the full symmetry of kagome.

We next establish that second-order interactions vanish for honeycomb links that share a site, e.g., $\langle 12 \rangle$ and $\langle 24 \rangle$ (nearest neighbors, distance 1/2 lattice spacing) shown in Fig.~\ref{sigma_x_interact}a. A second-order term $\sigma^x_{12}\sigma^x_{24}$ can be transformed into a first-order one, $\pm \sigma^x_{25}$, by using Gauss's law (\ref{constraint}). Because linear terms are forbidden, we conclude that $A^{(2)}_{12;24} = 0$.

By combining Gauss's law with lattice symmetries, we can rule out interactions between second neighbors (distance $\sqrt{3}/2$). In Fig.~\ref{sigma_x_interact}b, sites 2 and 5 carry opposite $Z_2$ charges, hence $\sigma^x_{21} \sigma^x_{24} \sigma^x_{25} = - \sigma^x_{53} \sigma^x_{56} \sigma^x_{52}$, or $\sigma^x_{21} \sigma^x_{24} = - \sigma^x_{53} \sigma^x_{56}$. Since $\sigma^x = \pm 1$, we obtain $\sigma^x_{21} \sigma^x_{53} = - \sigma^x_{24} \sigma^x_{56}$. By lattice symmetry, $A^{(2)}_{21;53} = A^{(2)}_{24;56}$. The two second-neighbor terms cancel out: 
\[
A^{(2)}_{21;53} \sigma^x_{21} \sigma^x_{53} + A^{(2)}_{24;56} \sigma^x_{24} \sigma^x_{56}
= A^{(2)}_{21;53} (\sigma^x_{21} \sigma^x_{53} + \sigma^x_{24} \sigma^x_{56}) = 0.
\]
Thus there are no interactions between second-neighbor links and we may set $A^{(2)}_{21;53} = A^{(2)}_{24;56} = 0$. 

The same line of argument shows that $A^{(2)}_{21;56} = A^{(2)}_{24;53} = 0$. This rules out a $\sigma^x \sigma^x$ term for links that are third neighbors (distance 1) residing on different hexagons of the honeycomb.

The closest non-vanishing interactions are between third neighbors (distance 1) residing on the same hexagon, e.g., links $24$ and $35$ in Fig.~\ref{sigma_x_interact}c. Adding these interactions takes us one step beyond the MSP model. 

To determine the sign of the coupling constant between these third neighbors, we use input from numerical studies. The DMRG calculations of \textcite{Yan} have identified a strong valence-bond resonance in the diamond-shaped loop, which corresponds to the tunneling between two quasi-degenerate dimer coverings (Fig.~\ref{sigma_x_interact}d). Translating the two configurations into electric flux variables (Fig.~\ref{sigma_x_interact}e), we see that electric fluxes $\sigma^x$ on opposite sides of a hexagon tend to be opposite, which indicates a negative coupling between the electric fluxes.

We have thus obtained a phenomenological Hamiltonian describing a $Z_2$ spin-liquid state of the kagome Heisenberg antiferromagnet:
\begin{align}
H&=-h\sum_{\alpha}F^z_{\alpha}+K\sum_{\mathrm{3.n.}}\sigma^x_{ij}\sigma^x_{kl}.
\label{pheno_hamil}
\end{align}
Pairwise interactions of electric fluxes $\sigma^x$ are limited to honeycomb links facing each other across a hexagon (Fig.~\ref{sigma_x_interact}c). Physical states satisfy Gauss-law constraints (\ref{constraint}). $K>0$, and $h>0$.

\section{Solving the model}\label{Method}

In this section, we present the techniques needed for studying the new model. We first map the $Z_2$ gauge theory (\ref{pheno_hamil}) to a quantum Ising model through a standard duality transformation. Details of the duality transformation are provided in Section \ref{duality_transform}. In Section \ref{softspin} we construct an analytically solvable soft-spin version of the dual Ising model. In Section \ref{observable} we translate $Z_2$ electric fluxes in the honeycomb lattice to the density of dimers on kagome. The latter is directly related to the nearest-neighbor spin correlations in the $S=1/2$ kagome Heisenberg antiferromagnet, connecting our phenomenological model to the DMRG results.

The dual Ising model is defined on a new, triangular lattice, whose sites are centers of honeycomb plaquettes, Fig.~\ref{fig_duality}a. We label the sites of the triangular lattice by Greek indices $\alpha, \beta, \gamma \ldots$ As usual, the magnetic-flux term translates into a transverse field coupled to new Ising variables $\tau^x_\alpha$. The interaction of electric fluxes translates into a standard Ising coupling $\tau^z_\alpha \tau^z_\beta$ between third neighbors of the new lattice. The system thus consists of four entirely decoupled sublattices (A, B, C, and D in Fig.~\ref{fig_duality}c) and behaves as four decoupled Ising models. 

\subsection{Duality transformation}\label{duality_transform}

\begin{figure}[tp]
\includegraphics[width=\columnwidth]{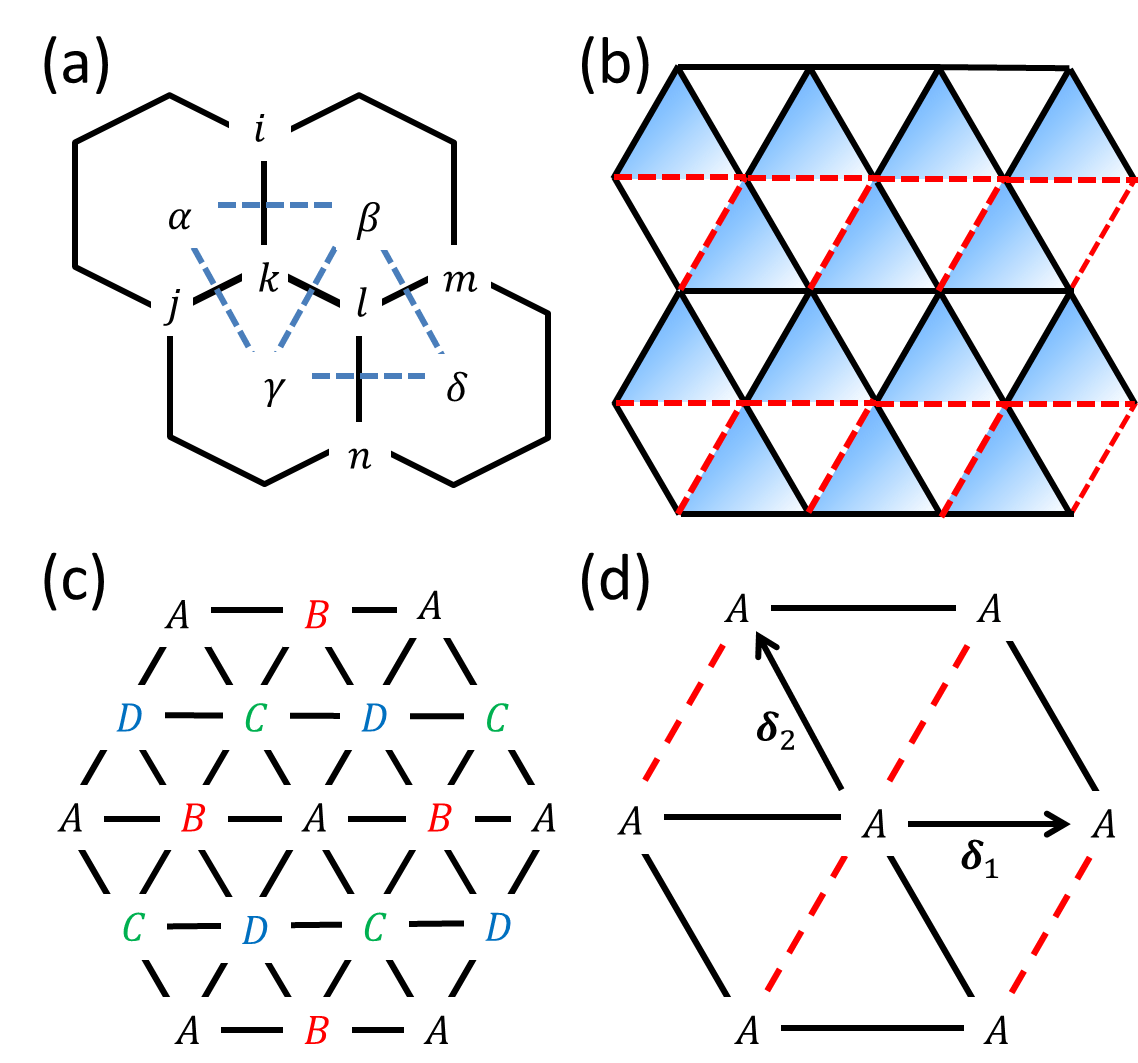}
\caption{(a) The dual Ising variables on defined on the hexagonal plaquettes, and they form a triangular lattice (blue dashed lines). Latin (Greek) letters are labels of the sites in the honeycomb (triangular) lattice. (b) The constraints on the phase factors $\lambda_{\alpha\beta}$. The product of the $\lambda_{\alpha\beta}$ around the shaded (unshaded) triangles is $-1$ ($1$). A possible arrangements for $\lambda_{\alpha\beta}$ is that $\lambda_{\alpha\beta}=1$ on black solid links and $-1$ on red dashed links. (c) The four sublattices of the triangular lattice. $A$, $B$, $C$, and $D$ are labels of the sublattices. Sites in the same sublattice are connected by the third-neighbor interaction. (d) The exchange interaction between the spins in A sublattice. The interaction is $K$ on black solid links and $-K$ on red dashed links. The primitive vectors $\bm{\delta}_{1,2}$ are also shown.}\label{fig_duality}
\end{figure}

The standard procedure to solve a $Z_2$ lattice gauge theory Hamiltonian is to map it to a quantum Ising model through a duality transformation. \cite{Savit} The electric flux operator $\sigma^x_{ij}$ is expressed in terms of new Ising variables $\tau^z_{\alpha}$ defined on honeycomb plaquettes: 
\begin{align}
\sigma^x_{ik}=\lambda_{\alpha\beta}\tau^z_\alpha\tau^z_\beta,
\label{duality1}
\end{align}
where $\alpha$ and $\beta$ are adjacent honeycomb plaquettes sharing honeycomb link $\langle{}ik\rangle$ (Fig.~\ref{fig_duality}a). $\lambda_{\alpha\beta}=\pm1$ is a $Z_2$ phase factor to be defined below. The magnetic flux operator translates into 
\begin{align}
F^z_{\alpha} = \tau^x_{\alpha}\label{duality2}.
\end{align}
Ising operators $\tau^x_\alpha$ and $\tau^z_\beta$ obey the standard anticommutation relation $\{\tau^x_{\alpha},\tau^z_{\beta}\}=\delta_{\alpha\beta}$. 

The phase factors $\lambda$ are required in the presence of nontrivial background charge. Substituting Eq.~\ref{duality1} into Gauss's law (\ref{constraint}) expresses the background charge on honeycomb site $\alpha\beta\gamma$ shared by the honeycomb plaquettes $\alpha$, $\beta$, and $\gamma$ to a product of $\lambda$ factors: \begin{equation}
\lambda_{\alpha\beta}\lambda_{\beta\gamma}\lambda_{\gamma\alpha} 
	=\left\{
		\begin{array}{cl}
			+1 & \textrm{down triangle},\\
			-1 & \textrm{up triangle}.
		\end{array}
	\right.{},
\label{lambda_condition}
\end{equation}
Fig.\ref{fig_duality}b shows a choice of $\lambda_{\alpha\beta}$ satisfying Eq.~\ref{lambda_condition}.

Equipped with the duality transformation, we are ready to map the $Z_2$ gauge theory Hamiltonian on the honeycomb lattice to a quantum Ising model on the dual triangular lattice. Substituting Eq.~\ref{duality1} and Eq.~\ref{duality2} into Eq.~\ref{pheno_hamil}, we obtain the dual Hamiltonian,
\begin{align}
H=-h\sum_{\alpha}\tau^x_\alpha+K\sum_{\langle\alpha,\beta\rangle\in\textrm{3.n.}}\lambda_{\alpha\gamma}\lambda_{\gamma\beta}\tau^z_\alpha\tau^z_\beta.\label{dual_hamil}
\end{align}
The second summation is over third-neighbor dual spin pairs. $\gamma$ is a triangular-lattice site located between third neighbors $\alpha$ and $\beta$. Since the interaction is only among third neighbors, the dual triangular lattice splits into four non-interacting sublattices A, B, C, and D (Fig.~\ref{fig_duality}c). The Hamiltonian is further simplified,
\begin{align}
H=H_A+H_B+H_C+H_D.\label{abcd}
\end{align}
The Hamiltonian for the A-sublattice is given by
\begin{align}
H_A=&-h\sum_{\bm{r}\in{}A}\tau^x_{\bm{r}}+K\sum_{\bm{r}\in{}A}\tau^z_{\bm{r}}\tau^z_{\bm{r}+\bm{\delta_1}}+K\sum_{\bm{r}\in{}A}\tau^z_{\bm{r}}\tau^z_{\bm{r}+\bm{\delta_2}}\nonumber\\
&-K\sum_{\bm{r}\in{}A}\tau^z_{\bm{r}}\tau^z_{\bm{r}+\bm{\delta_1+\delta_2}},\label{subhamil_a}
\end{align}
where $\bm{\delta}_{1,2}$ are primitive vectors generating the A-sublattice (Fig.~\ref{fig_duality}d). $H_{B,C,D}$ have the same form as $H_A$. Note that $H_A$ is unfrustrated because each triangle contains two antiferromagnetic links and one ferromagnetic link.

The decomposition of the dual triangular lattice into four sublattices is not accidental. It can be shown that the four sublattices remain decoupled if further-range two-spin interactions are included in the dual Ising model. However, the sublattices can be coupled by four-spin interactions. This feature of the dual theory is a manifestation of the $GL(2,\mathbb{Z}_3)$ projective symmetry identified by \textcite{Huh}, and we shall come back to this point in Section \ref{Discussion}.

The dual Ising model, being unfrustrated, possesses two phases at zero temperature. When $h\gg{}K$, the dual Ising model Eq.~\ref{subhamil_a} is in the paramagnetic phase, which corresponds to the $Z_2$ spin liquid phase of the quantum dimer model on kagome. When $h\ll{}K$, it is in the magnetic phase, which corresponds to the valence bond crystal phase of the quantum dimer model. \cite{Nikolic, Huh} We focus on the paramagnetic phase of the dual Ising model throughout this paper.

\subsection{Soft spin model}\label{softspin}

The dual Ising model Eq.~\ref{abcd} and Eq.~\ref{subhamil_a} cannot be easily solved. We construct an analytically solvable soft spin model that is expected to work well deeply in the disordered phase of the dual Ising model. \cite{J1J2} To this end, we replace the dual Ising operator $\tau^z_{\bm{r}} = \pm 1$ by a real-valued operator $\phi_{\bm{r}}$. The interaction between dual Ising spins therefore becomes
\begin{subequations}
\begin{equation}
K\sum_{\bm{r}\in{}A}(\phi_{\bm{r}}\phi_{\bm{r}+\bm{\delta_1}}+\phi_{\bm{r}}\phi_{\bm{r}+\bm{\delta_2}}-\phi_{\bm{r}}\phi_{\bm{r}+\bm{\delta_1+\delta_2}}).
\end{equation}
The transverse field term $-h\tau^x_{\bm{r}}$ brings quantum fluctuations to the dual Ising spin, so we add
\begin{equation}
\frac{1}{2}\sum_{\bm{r}\in{}A}\pi^2_{\bm{r}}+\frac{\Delta}{2}\sum_{\bm{r}\in{}A}\phi^2_{\bm{r}}
\end{equation}
\end{subequations}
to the Hamiltonian. $\pi_{\bm{r}}$ is the canonical momentum operator conjugate to $\phi_{\bm{r}}$. Here a mass term with $\Delta>0$ is introduced to ensure the fluctuations of $\phi_{\bm{r}}$ are gapped.

Combining all the terms listed above, we obtain the soft spin Hamiltonian for the A-sublattice,
\begin{align}
H^\textrm{soft}_A=&\frac{1}{2}\sum_{\bm{r}\in{}A}\pi^2_{\bm{r}}+\frac{\Delta}{2}\sum_{\bm{r}\in{}A}\phi^2_{\bm{r}}+K\sum_{\bm{r}\in{}A}\phi_{\bm{r}}\phi_{\bm{r}+\bm{\delta_1}}\nonumber\\
&+K\sum_{\bm{r}\in{}A}\phi_{\bm{r}}\phi_{\bm{r}+\bm{\delta_2}}-K\sum_{\bm{r}\in{}A}\phi_{\bm{r}}\phi_{\bm{r}+\bm{\delta_1+\delta_2}}\label{softspin_subhamil_a}
\end{align}
The soft spin Hamiltonian for sublattices $B$, $C$, and $D$ are obtained in the same way. The total soft spin Hamiltonian is 
\begin{align}
H^\textrm{soft}=H^\textrm{soft}_{A}+H^\textrm{soft}_{B}+H^\textrm{soft}_{C}+H^\textrm{soft}_{D}.\label{softspin_abcd}
\end{align}

\subsection{Dimer density}\label{observable}

\begin{figure}
\includegraphics[width=\columnwidth]{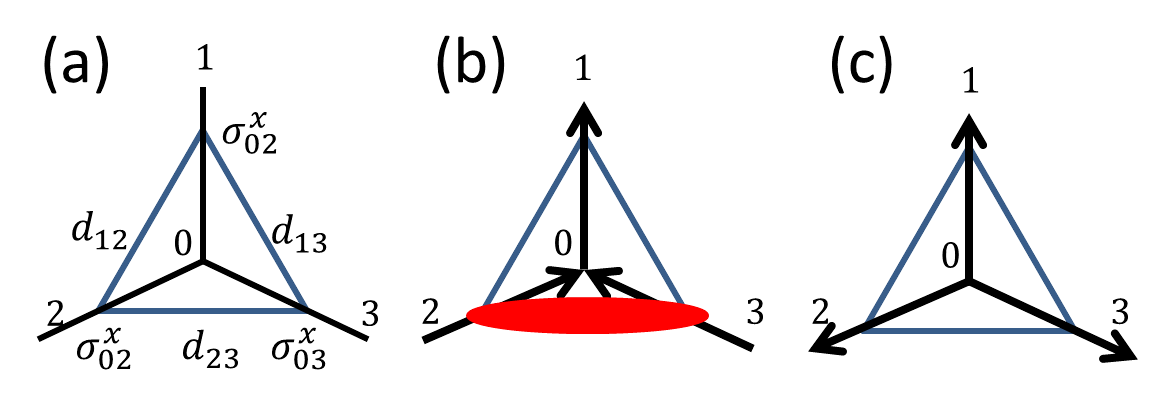}
\caption{(a) An up-triangle on kagome. $\sigma^x_{01}$, $\sigma^x_{02}$, and $\sigma^x_{03}$ are $Z_2$ electric flux operators on the links of the honeycomb lattice. $d_{12}$, $d_{13}$, and $d_{23}$ are the dimer density modulation on the links of kagome. (b) A triangle occupied by a dimer. (c) An empty triangle.}\label{dimer_density}
\end{figure}

The last piece of our construction is a formula relating the expectation value of electric flux operators to the density of dimers on kagome. To start with, we consider an up triangle on kagome, which corresponds to an A site in the honeycomb lattice (Fig.~\ref{dimer_density}a). The case for a down triangle or a B site is obtained by the same token. The dimer density modulation on kagome link $\langle{}ij\rangle$, $d_{ij}$, is defined as the dimer density on the said link minus the average number of dimers per link, $1/4$.

We postulate the following relation between the dimer density modulation and the expectation value of $Z_2$ electric flux operators on symmetry ground (Fig.~\ref{dimer_density}a),
\begin{align}
d_{23}=u\langle\sigma^x_{01}\rangle+v(\langle\sigma^x_{02}\rangle+\langle\sigma^x_{03}\rangle),\label{postulate}
\end{align}
where $u,v$ are coefficients to be determined. The expressions for $d_{12}$ and $d_{13}$ are obtained by $120^\circ$ rotations around the center of the triangle.

We consider the case in which one side of the triangle is occupied by a dimer and the corresponding arrow configuration (Fig.~\ref{dimer_density}b). On the one hand, the dimer density modulations are given by,
\begin{align}
d_{12} = -\frac{1}{4};\quad d_{13} = -\frac{1}{4};\quad d_{23} = \frac{3}{4}.
\end{align}
On the other hand, the expectation values of the $Z_2$ electric flux operators are given by
\begin{align}
\langle\sigma^x_{01}\rangle=1;\quad \langle\sigma^x_{02}\rangle=-1;\quad \langle\sigma^x_{03}\rangle=-1.
\end{align}
The postulated relation Eq.~\ref{postulate} is consistent with the above if and only if,
\begin{align}
u=\frac{1}{4};\quad{}v=-\frac{1}{4}.
\end{align}
Therefore, we obtain the following relation:
\begin{align}
d_{23}=\frac{1}{4}(\langle\sigma^x_{01}\rangle-\langle\sigma^x_{02}\rangle-\langle\sigma^x_{03}\rangle).\label{electric2dimer}
\end{align}

The self-consistency of Eq.~\ref{electric2dimer} can be further examined by considering another configuration shown in Fig.~\ref{dimer_density}c, where $d_{12} = d_{13} = d_{23} = -\frac{1}{4}$, and $\langle\sigma^x_{01}\rangle = \langle\sigma^x_{02}\rangle = \langle\sigma^x_{03}\rangle = 1$. It can be seen that Eq.~\ref{electric2dimer} holds in this special case as well.

\section{Results}\label{Results}

In Sections \ref{Model} and \ref{Method}, we have constructed a phenomenological $Z_2$ lattice gauge theory and developed necessary techniques to solve it deeply in the deconfined phase. In this section we compare predictions of our model with the DMRG results in various settings.

In Sections \ref{even_cylinder}, \ref{odd_cylinder}, and \ref{chiral_cylinder}, we study the ground state of the phenomenological model on kagome cylinders. We show that the ground state strongly depends on the cylindrical geometry. In Section \ref{dimerdimer}, we study the singlet-pinning effect in the ground state of kagome cylinders. In Section \ref{edgespinon}, we analyze the effect of the open boundary on the ground state of kagome cylinders and show that certain types of open boundary bind spinons near the edge. In all of the cases studied, we find good agreement between the phenomenological model and the DMRG.

\subsection{Ground state of YC$4m$ kagome cylinders}\label{even_cylinder}

Most of the recent DMRG studies on $S=1/2$ kagome Heisenberg antiferromagnet are performed with cylindrical geometry. \cite{Yan, Depenbrock, Jiang} The periodic boundary condition is imposed in one direction of kagome and open boundary condition in the other direction, making the system effectively a cylinder. The circumference of the cylinder is usually much smaller than the length, and we shall assume the length is infinite in the following discussion.

The DMRG study has identified an even-odd effect.\cite{Yan, White} A kagome cylinder can be regarded as a one-dimensional system, and the unit cell is defined as the smallest building block that generates the whole cylinder by translation in the length direction. The ground state is a uniform spin liquid when the number of spins per unit cell is even. When the number is odd, the spin-liquid ground state coexists with a valence-bond density wave pattern, which breaks the spatial symmetries of the cylinder.

\begin{figure}
\includegraphics[width=0.9\columnwidth]{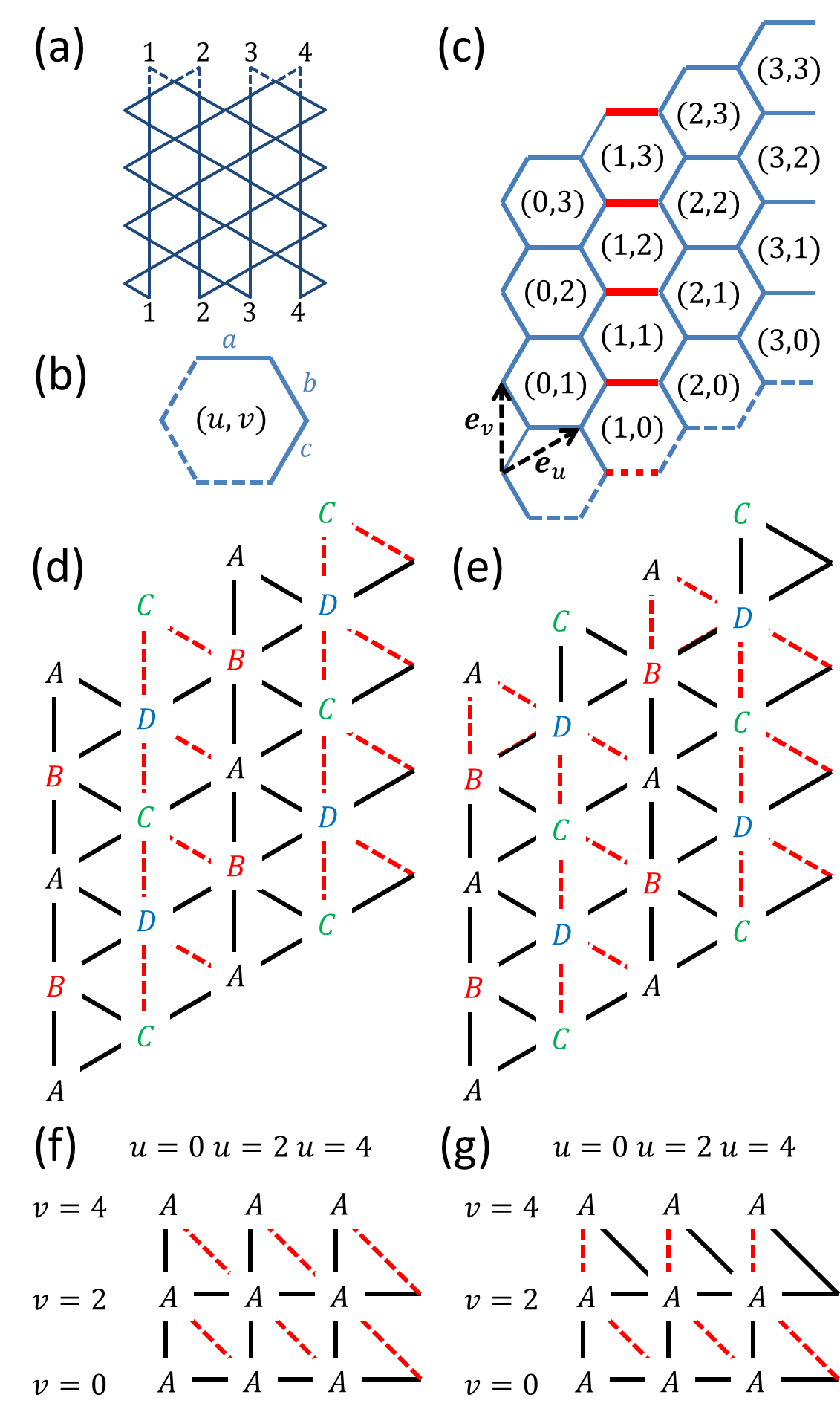}
\caption{(a) The YC8 cylinder. The periodic boundary condition is imposed along the vertical direction. Sites with the same number are identical. (b) The labeling system employed in the main text. (c) The corresponding $Z_2$ gauge theory model on the honeycomb lattice. Arrows are the primitive vectors $\mathbf{e}_{u}$ and $\mathbf{e}_{v}$ of the dual triangular lattices. The coordinates of the dual triangular sites are shown. (d) The dual model for YC8 cylinder in the even sector. The phase factor $\lambda=1(-1)$ on black solid (red dashed) links. (e) The same dual model in the odd sector. (f) The exchange interaction between dual spins in A sublattice in the even sector. The interaction is $K$ on black solid links and $-K$ on red dashed links. (g) The exchange interaction between dual spins in A sublattice in the odd sector.}\label{yc8}
\end{figure}

Motivated by the DMRG findings, we study the phenomenological model on cylinders. To start with, we consider the so-called ``YC$4m$'' cylinders, where $m\in\mathbb{Z}$ and the number $4m$ denotes the circumference of the tube. \cite{Yan} Note that the number of spins per unit cell is $6m$, which is even. Fig.~\ref{yc8}a shows a YC8 cylinder. We shall use the YC8 cylinder as an illustrative example in the following discussion, and all the results can be easily generalized to all YC$4m$ cylinders.

The corresponding $Z_2$ gauge theory model is defined on the honeycomb lattice with ``armchair'' type periodic boundary condition (Fig.~\ref{yc8}c). Note that the honeycomb cylinder is unfolded in a way different from Fig.~\ref{yc8}a.

We introduce a labeling system that shall be employed in later discussions. The honeycomb plaquettes are generated by primitive vectors $\mathbf{e}_u$ and $\mathbf{e}_v$. The position vector of a honeycomb plaquette (or a dual triangular site) is uniquely expressed as $u\mathbf{e}_u+v\mathbf{e}_v$, and it is labeled as $(u,v)$. Then, the three honeycomb bonds adjacent to the said plaquette are labeled as $(u,v,a)$, $(u,v,b)$, and (u,v,c) (Fig.~\ref{yc8}b).
 
The $Z_2$ gauge theory Hamiltonian possesses a global conservation law associated with the cylindrical topology. \cite{Kitaev} We consider a non-contractible contour around the cylinder, shown as thick red bonds in Fig.~\ref{yc8}c. We define the total electric flux passing the contour,
\begin{align}
X=\prod_{v}\sigma^x_{1,v,a}.\label{xdef}
\end{align}
It can be shown that $X$ commutes with the Hamiltonian $[X,H]=0$, meaning the total $Z_2$ electric flux is conserved. Therefore, the Hilbert space is divided into two topological sectors, with $X=1$ and $X=-1$ respectively.

The dual Ising model is defined on triangular lattice with cylindrical geometry (Figs.~\ref{yc8}d,e). The global conservation law on total $Z_2$ electric flux $X$ gives rise to new constraints on the phase factors $\lambda_{\alpha\beta}$ in addition to Eq.~\ref{lambda_condition}. Substituting Eq.~\ref{duality1} into Eq.~\ref{xdef}, we rewrite the total $Z_2$ electric flux operator $X$ in terms of dual variables (Fig.\ref{yc8}b):
\begin{align}
X=\prod_{v}\lambda_{(1,v),(1,v+1)}=\pm{}1\label{lambda_condition2}
\end{align}
Thus, the total electric flux $X$ is dual to the total flux of $\lambda_{\alpha\beta}$ around the cylinder circumference.

In what follows, we treat the even ($X=1$) and odd ($X=-1$) sectors separately. We consider the even sector at first. A choice of $\lambda_{\alpha\beta}$ satisfying conditions Eq.~\ref{lambda_condition} and Eq.~\ref{lambda_condition2} is shown in Fig.~\ref{yc8}c. We see that the dual model is decomposed into four independent copies of unfrustrated quantum Ising model, corresponding to four sublattices of the triangular lattice. In the disordered phase of the dual model, or equivalently the deconfined phase of the $Z_2$ gauge theory model,
\begin{align}
\langle\sigma^x_{ij}\rangle=\lambda_{\alpha\beta}\langle\tau^z_{\alpha}\tau^z_{\beta}\rangle=\lambda_{\alpha\beta}\langle\tau^z_{\alpha}\rangle\langle\tau^z_{\beta}\rangle=0.
\end{align}
The second equality follows from the fact that dual spins $\alpha$ and $\beta$, being nearest neighbors to each other, must belong to two different sublattices. The third equation follows from the assumption that the dual model is in the disordered phase. Given the modulation of dimer density is proportional to $\langle\sigma^x_{ij}\rangle$, we deduce that the ground state is uniform for YC$4m$ cylinders in the even sector.

We proceed to calculate the ground state energy of the YC$4m$ cylinder in the even sector. The circumference of each sublattice is $m$. The soft spin model for the A sublattice is given by (Fig.~\ref{yc8}f)
\begin{align}
H^\textrm{soft}_A&=\frac{1}{2}\sum_{(u,v)\in{}A}\pi^2_{u,v}+\frac{\Delta}{2}\sum_{(u,v)\in{}A}\phi^2_{u,v}+K\sum_{u,v}(\phi_{u,v}\phi_{u+2,v}\nonumber\\
&+\phi_{u,v}\phi_{u,v+2}-\phi_{u,v}\phi_{u-2,v+2}),\label{yc4m_soft_spin}
\end{align}
Here the periodic boundary condition $\phi_{u,v}=\phi_{u,v+2m}$ is imposed. The above Hamiltonian can be readily diagonalized. The ground state energy per dual site is given by
\begin{align}
\epsilon_{1}(m)=\frac{1}{2m}\sum^{m-1}_{v=0}\int^{2\pi}_{0}\frac{dk_u}{2\pi}\omega\left(k_u,\frac{2\pi{}v}{m}\right),\label{gs_energy_even}
\end{align}
where
\begin{equation}
\omega=\sqrt{\Delta+2K[\cos{k_u} + \cos{k_v}
-\cos(k_u-k_v)]}
\label{dispersion}
\end{equation}
is the dispersion relation of soft spin fluctuations.

The analysis on the dual Ising model in the odd sector can be carried out in the similar manner. A choice of $\lambda_{\alpha\beta}$ is shown in Fig.~\ref{yc8}e. Similar to the even sector, the dual triangular lattice is decomposed into four independent, identical sublattices, and the ground state is uniform in the odd sector as well.

As shown in Fig.~\ref{yc8}g, the explicit translational invariance of the sublattice Hamiltonian is lost in the circumference direction. However, we can make a gauge transformation $\phi_{u,2m}\to{}-\phi_{u,2m}$ to restore the explicit translational invariance. The soft spin model becomes identical to Eq.~\ref{yc4m_soft_spin}. The boundary condition, however, becomes anti-periodic: $\phi_{u,v}=-\phi_{u,v+2m}$. The ground state energy per site is given by
\begin{align}
\epsilon_{-1}(m)=\frac{1}{2m}\sum^{m-1}_{v=0}\int^{2\pi}_{0}\frac{dk_u}{2\pi}\omega\left(k_u,\frac{(2v+1)\pi}{m}\right),\label{gs_energy_odd}
\end{align}
Here $\omega$ is the same as Eq.~\ref{dispersion}.

\begin{figure}
\includegraphics[width=\columnwidth]{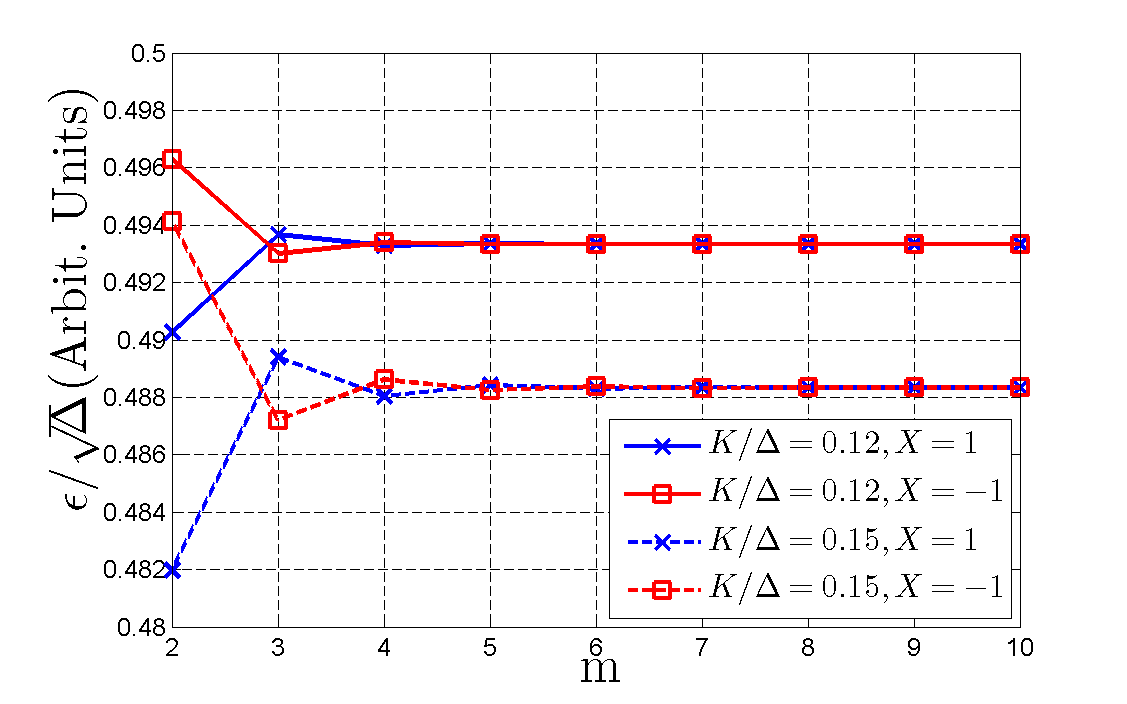}
\caption{Ground state energy per site $\epsilon$ of YC$4m$ cylinder in even $X=1$ (blue crosses) and odd $X=-1$ sectors (red open circles). The solid and dashed lines correspond to $K/\Delta=0.12$ and $K/\Delta=0.15$, respectively.}\label{yc4m_energy}
\end{figure}

Fig.\ref{yc4m_energy} shows the ground state energy per dual site as a function of cylinder circumference for generic values of $\Delta$ and $K$, based on numerical evaluation of Eq.~\ref{gs_energy_even} and Eq.~\ref{gs_energy_odd}. We find that the ground state energy in the two sectors becomes degenerate in the limit of infinite circumference as expected for a gapped $Z_2$ spin liquid. When the circumference is finite, the energy splitting shows an alternating pattern: the even ground state has lower energy when $m$ is even, and the odd ground state has lower energy otherwise. Such a pattern plays an important role in our discussion on edge spinons in Section \ref{edgespinon}.

To summarize, we have found that the ground state of the YC$4m$ cylinder is the uniform $Z_2$ spin liquid. The ground state is in the even sector $X=1$ for even $m$, and the odd sector $X=-1$ for odd $m$.

\subsection{Ground state of YC$4m+2$ kagome cylinders}\label{odd_cylinder}

\begin{figure}
\includegraphics[width=0.9\columnwidth]{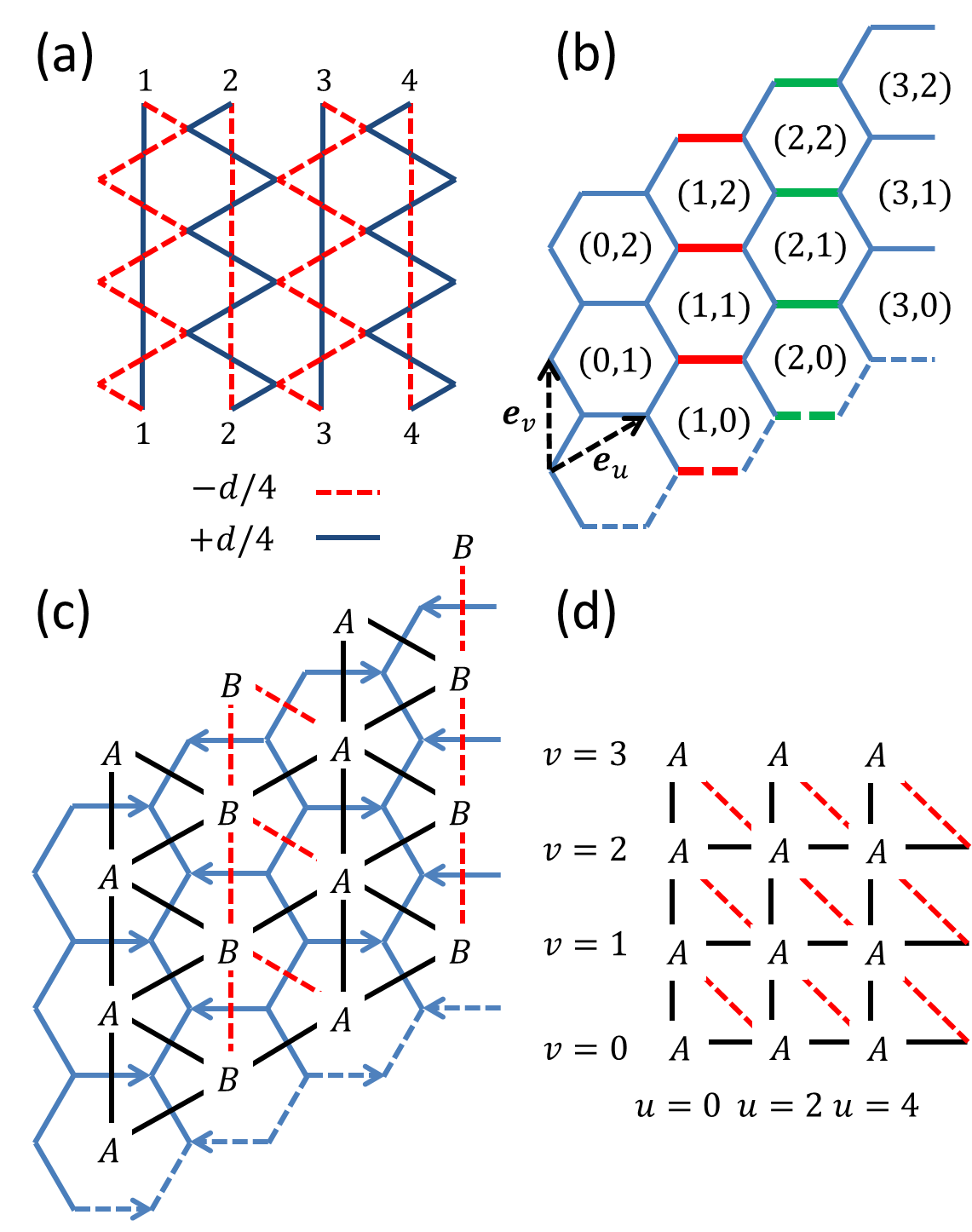}
\caption{(a) A YC6 cylinder and the predicted valence-bond density modulation pattern. Lattice sites with the same numeral label are identical. (b) The corresponding $Z_2$ gauge theory. (c) The dual Ising model. A and B are sublattice labels. $\lambda=1(-1)$ on black solid (red dashed) links. (d) The interaction between $A$ dual spins. The interaction is $K(-K)$ on black solid (red dashed) links.}\label{yc6}
\end{figure}

In this section, we discuss another family of kagome cylinders known as YC$4m+2$, $m\in\mathbb{Z}$. \cite{Yan} The number of spins per unit cell is $6m+3$, which is odd. Based on the Lieb-Schultz-Mattis theorem, we expect that the ground state is either gapless or symmetry-breaking. \cite{Lieb} Indeed, DMRG has identified a symmetry-breaking valence-bond density wave pattern. We will show that our model reproduces the same pattern in the ground state.

As a concrete example, we consider the YC6 cylinder, a member of the YC$4m+2$ family, in the following discussion. The analysis is carried on in parallel with Section \ref{even_cylinder}. Fig.~\ref{yc6}a shows the YC6 kagome cylinder, and Fig.~\ref{yc6}b shows its corresponding honeycomb cylinder on which the $Z_2$ gauge theory is defined. The total $Z_2$ electric flux in the $Z_2$ gauge theory model is defined as
\begin{align}
X=\prod_{v}\sigma^x_{1,v,a},
\end{align}
corresponding to the contour highlighted as the thick red bonds in Fig.~\ref{yc6}a. $[X,H]=0$, and the Hilbert space falls into two topological sectors, $X=1$ (even) and $X=-1$ (odd).

The two topological sectors are related by translational operation. To see this, we consider another contour shown as the thick green bonds in Fig.~\ref{yc6}b. The associated total electric flux operator is defined as:
\begin{align}
X'=\prod_{v}\sigma^x_{2,v,a}.
\end{align}
On the one hand, it can be seen that $X'=-X$ on account of Gauss's law (\ref{constraint}). On the other hand, the new contour is related to the old one by a $\mathbf{e}_u$ shift, and we have $T(\mathbf{e}_u)XT(\mathbf{e}_u)^{\dagger}=X'$. Here $T(\mathbf{e}_u)$ denotes the associated shift operator. Therefore,
\begin{align}
T(\mathbf{e}_u)X=-XT(\mathbf{e}_u).\label{TX_anti_commute}
\end{align}

Eq.~\ref{TX_anti_commute} implies that the two topological sectors are degenerate in energy in the YC6 cylinder. Note $T(\mathbf{e}_u)$ is a symmetry,
\begin{align}
[T(\mathbf{e}_u),\,{}H]=0.
\end{align}
Let $|\psi\rangle$ be a common eigenstate of the Hamiltonian $H$ and the total electric flux $X$:
\begin{align}
H|\psi\rangle=\epsilon|\psi\rangle,\quad{}X|\psi\rangle=x|\psi\rangle.
\end{align}
Then,
\begin{align}
HT(\mathbf{e}_u)|\psi\rangle&=T(\mathbf{e}_u)H|\psi\rangle=\epsilon{}T(\mathbf{e}_u)|\psi\rangle,\nonumber\\
\quad{}XT(\mathbf{e}_u)|\psi\rangle&=-T(\mathbf{e}_u)X|\psi\rangle=-xT(\mathbf{e}_u)|\psi\rangle.
\end{align}
We see $|\psi\rangle$ and $T(\mathbf{e}_u)|\psi\rangle$ are degenerate in energy yet belong to two different topological sectors. Therefore, the energy spectra in the two topological sectors coincide, and the degenerate eigenstates are related by the $\mathbf{e}_u$ translation. Most importantly, in contrast to the YC$4m$ cylinders, the ground states in the two topological sectors of the YC$4m+2$ cylinders are exactly degenerate, and they break the translation symmetry in $\mathbf{e}_u$ direction.

We proceed to discuss the dual soft spin model. We shall consider the even sector only (Fig.~\ref{yc6}c). The circumference of the dual triangular lattice is $2m+1$, making it impossible to partition the lattice into four independent sublattices. Consequently, there are only two independent sublattices, labeled as A and B. The Hamiltonians for both sublattices are identical. The soft spin model for the A sublattice is given by (Fig.~\ref{yc6}d):
\begin{align}
H^\textrm{soft}_A&=\frac{1}{2}\sum_{(u,v)\in{}A}\pi^2_{u,v}+\frac{\Delta}{2}\sum_{u,v}\phi^2_{u,v}+K\sum_{u,v}(\phi_{u,v}\phi_{u+2,v}\nonumber\\
&+\phi_{u,v}\phi_{u,v+2}-\phi_{u,v}\phi_{u-2,v+2}),
\end{align}
The periodic boundary condition $\phi_{u,0}=\phi_{u,2m+1}$ is imposed. 

We have deduced on symmetry ground that the ground state in the even sector must break the translational symmetry. In what follows, we show that the translation symmetry is broken by a dimer-density modulation pattern. To this end, we calculate the expectation value of $Z_2$ electric flux in the ground state. For the $(u,v,b)$ bonds,
\begin{align}
\langle\sigma^x_{u,v,b}\rangle=\lambda_{(u,v),(u+1,v)}\langle\tau^z_{u,v}\tau^z_{u+1,v}\rangle\nonumber\\
=\lambda_{(u,v),(u+1,v)}\langle\tau^z_{u,v}\rangle\langle\tau^z_{u+1,v}\rangle=0.\label{elec_flux_b_bonds}
\end{align}
The second equality follows from the fact that the dual sites $(u,v)$ and $(u+1,v)$ belong to different sublattices and therefore are uncorrelated (Fig.~\ref{yc6}c). The third equality follows from the assumption that the dual model is in the disordered phase. By the same token,
\begin{align}
\langle\sigma^x_{u,v,c}\rangle=0.\label{elec_flux_c_bonds}
\end{align}

The expectation value of $\sigma^x_{u,v,a}$ does not vanish in the ground state.
\begin{align}
\langle\sigma^x_{u,v,a}\rangle&\approx\lambda_{(u,v),(u,v+1)}\langle\phi_{u,v}\phi_{u,v+1}\rangle\\
&=\left\{
\begin{array}{cc}
+d & \mbox{ if } u = {}0\,\textrm{mod}\,2,\\
-d & \mbox{ if } u = {}1\,\textrm{mod}\,2.
\end{array}
\right.\label{elec_flux_a_bonds}
\end{align}
Here $d\neq0$ in general because the dual sites $(u,v)$ and $(u,v+1)$ belong to the same sublattice (Fig.~\ref{yc6}c). The plus and minus signs come from the $\lambda$ phase factor. We have defined
\begin{align}
d\equiv\langle\phi_{u,v}\phi_{u,v+1}\rangle,
\end{align}
where $d$ doesn't depend on $u,v$ thanks to the translational invariance of the sublattice Hamiltonian and the fact that the Hamiltonians for A and B sublattices are identical.

Combing Eq.~\ref{elec_flux_b_bonds}, Eq.~\ref{elec_flux_c_bonds}, and Eq.~\ref{elec_flux_a_bonds}, we conclude that the $Z_2$ lattice gauge theory displays an alternating electric-flux pattern in the ground state on the YC$4m+2$ cylinders. The pattern in the even sector is shown in Fig.~\ref{yc6}c as arrows. The pattern in the odd sector is obtained by a $\mathbf{e}_u$ shift, or equivalently by reversing all the arrows.

The $Z_2$ electric flux operator is tied to the density modulation. We can therefore determine the dimer density modulation in the ground state of YC$4m+2$ cylinders by using the results in Section \ref{observable}. The dimer modulation pattern is shown in Fig.~\ref{yc6}a. (The pattern in the other sector is obtained by a shift of $\mathbf{e}_u$.) The blue solid and red dashed bonds stand for dimer density modulation of $d/4$ and $-d/4$ respectively. In terms of the Heisenberg model, they correspond to stronger and weaker nearest-neighbor spin-spin correlation. The pattern shown in Fig.~\ref{yc6}a agrees well with the DMRG result. \cite{White}

\begin{figure}
\includegraphics[width=\columnwidth]{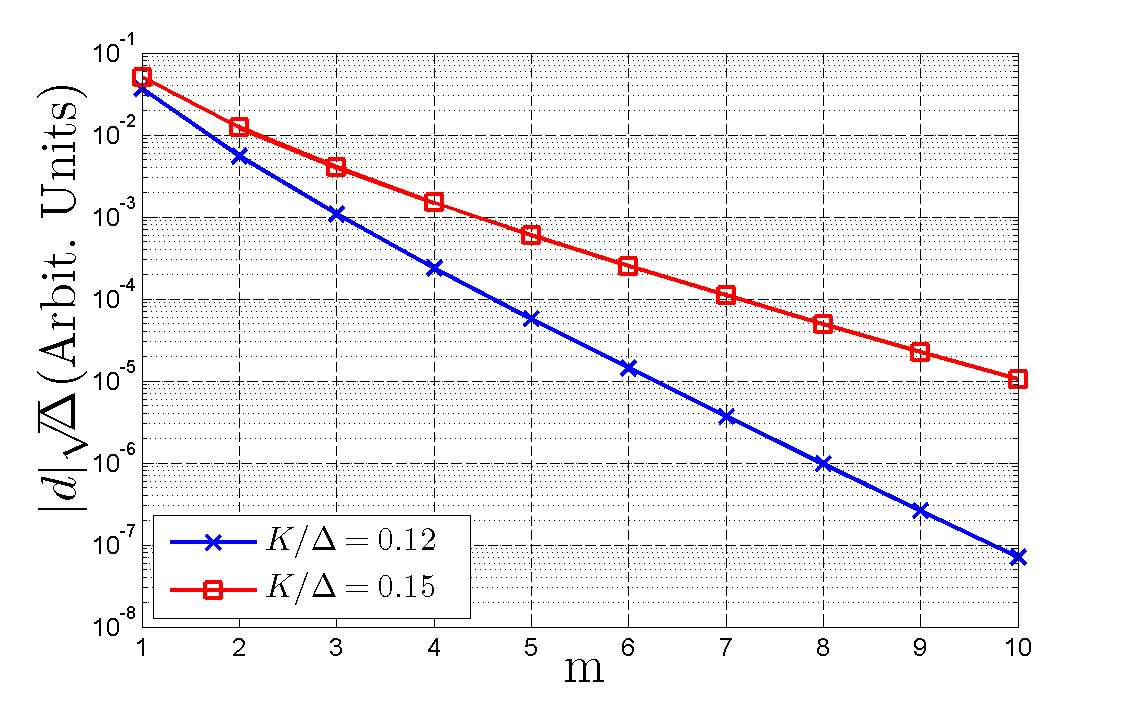}
\caption{Dimer density modulation $|d|$ in the ground states of YC$4m+2$ cylinders as a function of the cylinder index $m$. }\label{yc4mp2_mod}
\end{figure}

The value of $d$ can be explicitly calculated by using the soft spin model. We obtain,
\begin{align}
d=\frac{1}{2(2m+1)}\sum^{2m}_{v=0}\int^{2\pi}_0\frac{dk_u}{2\pi}\frac{\cos[2\pi{}v/(2m+1)]}{\omega[k_u,2\pi{}v/(2m+1)]},
\end{align}
where
\begin{equation} \omega = \sqrt{\Delta+2K[\cos{k_u}+\cos{2k_v}
-\cos(k_u+2k_v)]}.
\end{equation}
Fig.~\ref{yc4mp2_mod} shows the absolute value of $d$ as a function of cylinder index $m$ for generic value of model parameters. We can see that $|d|$ decreases almost exponentially as $m$ increases, reflecting the finite correlation length in the ground state.

To sum up, we have found that the ground states of the YC$4m+2$ cylinders are two-fold degenerate. The ground states display a  symmetry-breaking valence-bond modulation pattern, which agrees with what found by DMRG.

\subsection{Ground states of YC$(4m+1)$-2 kagome cylinders}\label{chiral_cylinder}

\begin{figure}
\includegraphics[width=0.9\columnwidth]{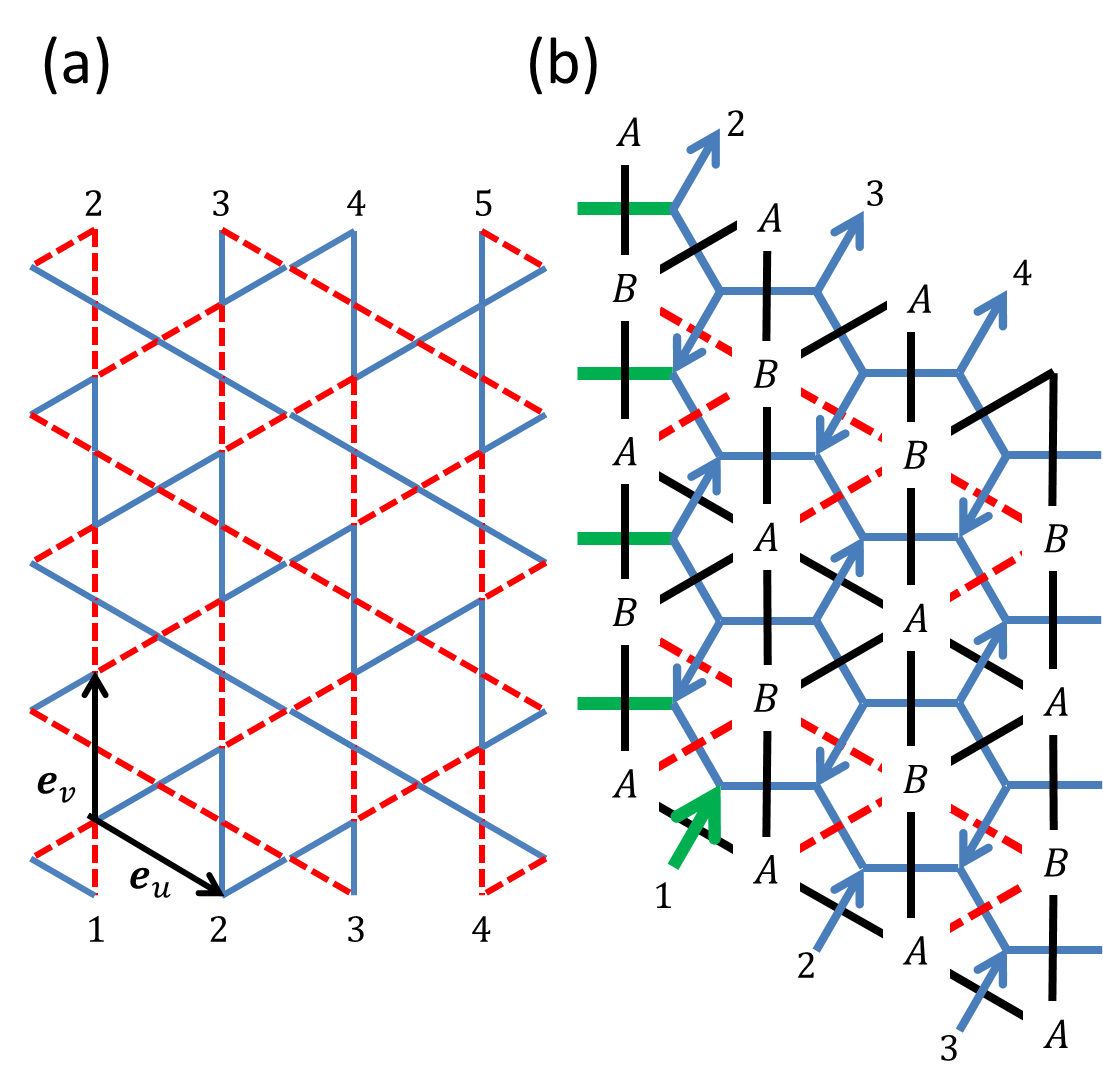}
\caption{(a) Valence-bond modulation pattern on the YC9-2 cylinder calculated from the phenomenological theory. The spin-spin correlation is enhanced on the solid blue bonds and weakened on the dashed red bonds, respectively. Sites labeled with the same number are identical. (b) The corresponding gauge theory model and the dual Ising model. Arrows indicate the non-vanishing expectation value of $\sigma^x_{ij}$ in the ground state. Boundary bonds labeled with the same number are identical. Thick green bonds highlight the contour with respect to which the total electric flux operator $X$ is defined. $A$ and $B$ stand for two independent sublattice sites of the dual lattice. The phase factor $\lambda=1(-1)$ on black solid (red dashed) bonds. Here the even sector $X=1$ is shown.}\label{yc9-2}
\end{figure}

In this section, we briefly discuss one more family of kagome cylinders dubbed YC$(4m+1)$-2 \cite{Yan}. This new family of cylinders lacks for the reflection symmetry along the circumference direction. The DMRG calculation has identified an valence-bond modulation pattern similar to the one found in the YC$4m+2$ family but with a different orientation. The analysis on YC$(4m+1)$-2 cylinders can be carried out in the same manner as in Section \ref{odd_cylinder}, and therefore we simply summarize the result.

Fig.~\ref{yc9-2}a shows the YC9-2 cyliner, a member of the YC$(4m+1)$-2 family. Note that the periodic boundary condition along the circumference direction of cylinder is accompanied by a shift in the length direction. Fig.~\ref{yc9-2}b shows the corresponding gauge theory model and the dual Ising model. The topological sectors are defined with respect to the contour highlighted as the thick green bonds. Similar to the YC$4m+2$ cylinders, the energy spectra in both sectors are exactly identical. In particular, the ground states are two-fold degenerate, and they break the translation symmetry along the $\mathbf{e}_v$ direction. Here the expectation value of $\sigma^x_{ij}$ in the even-sector ground state is shown as arrows, from which we determine the valence-bond modulation pattern (Fig.~\ref{yc9-2}a). The pattern in the other ground state is obtained by a $\mathbf{e}_v$ translation.

Comparing the calculated pattern with the one obtained by DMRG, we find that the two agree qualitatively. However, the pattern predicted by our phenomenological theory is perfectly uniform along the $\mathbf{e}_u$ direction, whereas the DMRG pattern shows another, weaker, modulation along the $\mathbf{e}_u$ direction \cite{Yan}. This discrepancy may be due to the shorter-range physics that is not included in our theory.

\subsection{Singlet-pinning effects in the ground state}\label{dimerdimer}

\begin{figure}
\includegraphics[width=\columnwidth]{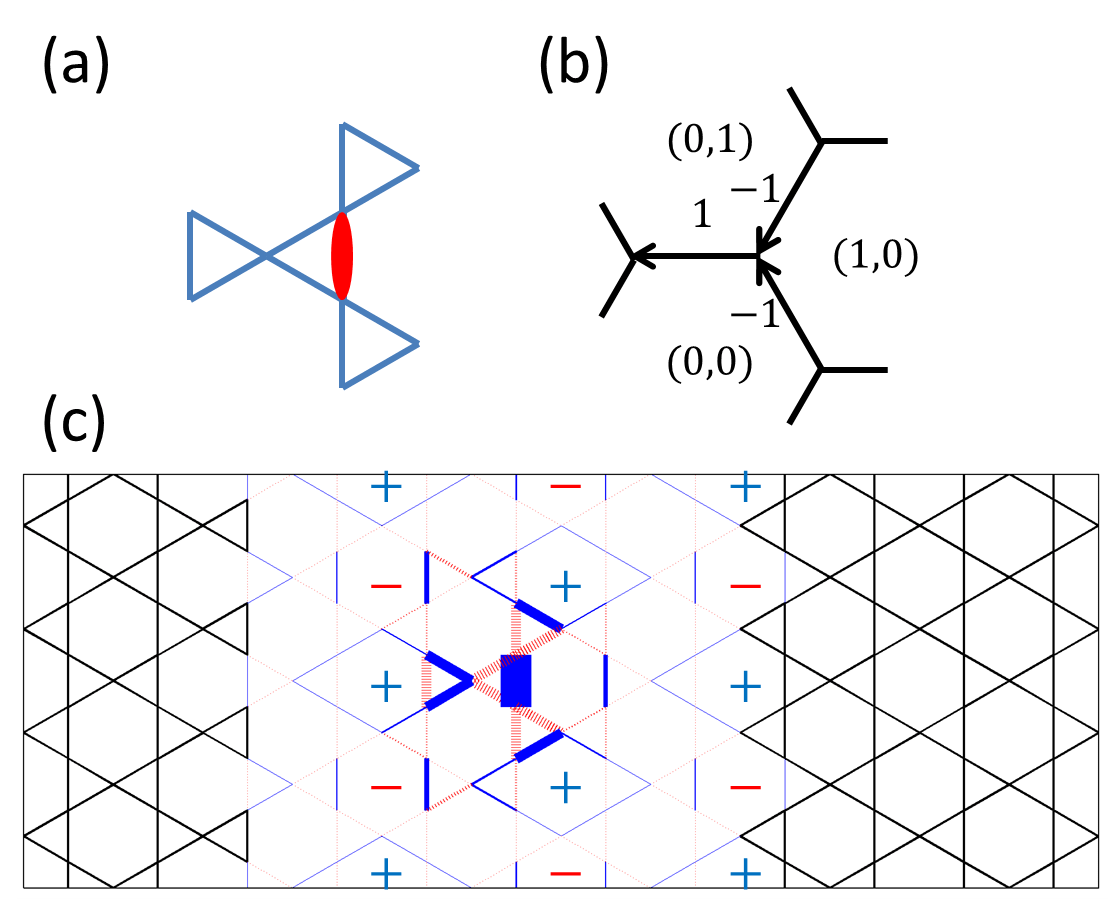}
\caption{(a) A singlet or dimer is pinned on a bond with enhanced Heisenberg exchange interaction. (b) A pinned dimer is amount to frozen arrows in the arrow representation. (c) The dimer density modulation pattern on YC8 cylinder calculated from the phenomenological theory. We choose $\Delta=1$, $K=0.15$, and $V=0.2$. The dimer density is increased on blue solid bonds and decreased on red dashed bonds. The thickness is proportional to the magnitude of modulation.}\label{dimer_pinning}
\end{figure}

In this section, we study the response of the $Z_2$ spin liquid ground state to an externally-pinned singlet. Specifically, when the Heisenberg exchange interaction on a bond is enhanced by amount of $\delta{}J$, it is energetically favorable to form a spin singlet on that bond (Fig.~\ref{dimer_pinning}a). The pinned singlet freezes the resonating singlets nearby, and a modulation pattern of nearest-neighbor spin-spin correlation $\langle\mathbf{S}_i\cdot\mathbf{S}_j\rangle$ appears in its neighborhood. When $\delta{}J\ll{}1$, the pattern coincides with the dimer-dimer correlation in the unperturbed ground state. The goal of this section is to compute such a modulation pattern by using the phenomenological model and to compare it with the numerical results.

The effects of an enhanced bond in the Heisenberg model are effectively described by an dimer-attracting potential in the quantum dimer model, which is amount to introducing favored arrow orientations in the arrow representation (Fig.\ref{dimer_pinning}b). The $Z_2$ gauge theory Hamiltonian is given by
\begin{align}
H=H_0-V(\sigma^x_{0,0,a}-\sigma^x_{0,0,b}-\sigma^x_{0,1,c}),\label{dimer_pin_hamil}
\end{align}
where $H_0$ is the unperturbed Hamiltonian and $V$ is the strength of the pinning potential. The above Hamiltonian is readily solved by using the duality mapping and the soft spin approximation. 

Fig.~\ref{dimer_pinning}c shows the calculated dimer density modulation pattern on the YC8 cylinder by solving Eq.~\ref{dimer_pin_hamil} for $\Delta=1$, $K=0.15$, and $V=0.2$. Note that the unperturbed ground state ($V=0$) of the YC8 cylinder is uniform as discussed in Section.\ref{even_cylinder}. The perturbed Hamiltonian is solved in the even sector, where the ground state of the YC8 cylinder lies (Fig.\ref{yc4m_energy}). The dimer density is increased on the blue solid bonds and decreased on the red dashed bonds. The bond thickness is proportional to the magnitude of modulation. The bond with enhanced exchange interaction has the largest dimer density.

We find qualitative agreement between the phenomenological theory and the DMRG. \cite{White} In addition, we note the following features of the modulation pattern calculated from the phenomenological theory. Firstly, the spatial modulation quickly decays as the distance to the enhanced bond increases, which is another manifestation of the gapped nature of the ground state. Secondly, there are alternatively enhanced and reduced diamond shapes which tile the whole cylinder, highlighted as plus and minus signs in Fig.\ref{dimer_pinning}c. These diamond shapes coincide with the so-called ``diamond resonance'' observed by DMRG \cite {White} and Lanczos diagonalization studies \cite{Lauchli}.

\subsection{Edge spinons and $Z_2$ screening}\label{edgespinon}

\begin{figure}
\includegraphics[width=\columnwidth]{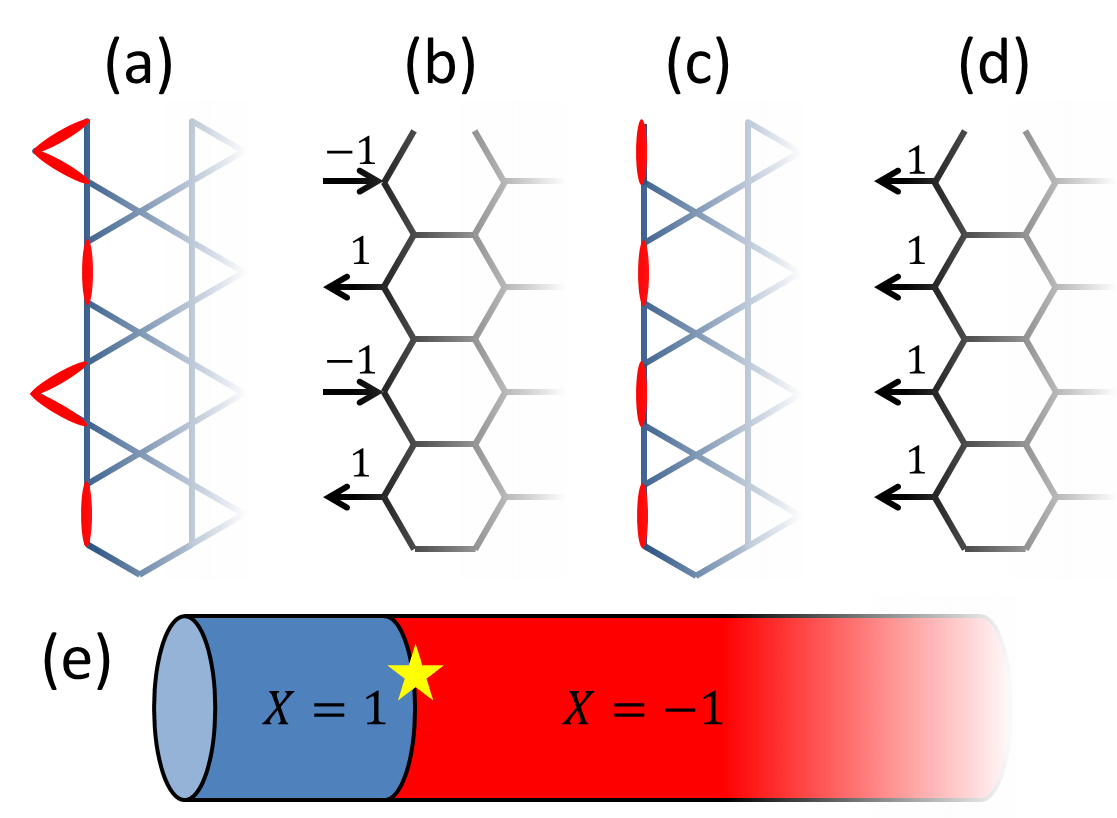}
\caption{(a) The open boundary of YC$4m$ cylinders employed by \textcite{Yan}. The singlets are pinned at the boundary. (b) The arrow representation for the frozen dimer covering shown in (a). The numbers $\pm1$ are the $Z_2$ electric fluxes on the dangling bonds. (c) Another open boundary condition of YC$4m$ cylinders and the pinned singlets. (d) The arrow representation and the $Z_2$ electric fluxes corresponding to the frozen dimer covering shown in (c). (e) The electric flux in the bulk is screened by a spinon.}\label{edge}
\end{figure}

In our previous discussion, we assume that the kagome cylinders are infinite in the length direction. As we shall see below, the open boundaries give rise to a new phenomenon: in certain circumstances, an open boundary can bind a spinon nearby. The edge spinons are detectable in numerical simulation, which could serve as a direct evidence for fractionalization. \cite{Jiang2}

To start with, we consider a particular type of open boundary of YC$4m$ cylinders, which is employed by \textcite{Yan} (Fig.~\ref{edge}a). The motivation is to make the open boundary compatible with the ``diamond resonance'' pattern. It is energetically favorable to form singlets on dangling bonds, and therefore a particular dimer covering is pinned at the boundary, which has been observed in DMRG calculations.

The physical meaning of such an open boundary condition becomes transparent in the $Z_2$ gauge theory description. Converting the frozen dimer covering to arrow representation, we find that the effect of the open boundary is to enforce an alternating $Z_2$ electric flux pattern at the edge (Fig.~\ref{edge}b). The total edge electric flux is $(-1)^m$ in the YC$4m$ cylinder. As discussed in Section \ref{odd_cylinder}, the total electric flux in the ground state of YC$4m$ cylinder is $(-1)^m$. Comparing the total electric flux enforced by the open boundary condition with the one selected by the bulk energetics, we find that the two match for any $m$.

However, the agreement between the edge total electric flux and the bulk total electric flux can be lost if one chooses other types of open boundary. An example is shown in Fig.~\ref{edge}c. Similarly, dimers are pinned at the boundary, and an electric flux pattern is enforced at the edge (Fig.~\ref{edge}d). The total electric flux at the edge is always $1$ for any YC$4m$ cylinder. When $m$ is odd, the edge total electric flux and the bulk total electric flux don't match.

The mismatching is resolved by a spinon screening the bulk electric flux in the ground state of YC$4m$ cylinder when $m$ is odd, which is in close analogy to the screening phenomenon in a metal (Fig.~\ref{edge}e). Close to the edge, the total electric flux is $1$, and the cylinder segment between the edge and the spinon is in the even sector. Deep into the bulk, the total electric flux is $-1$, and the bulk is in the odd sector. Assuming that the distance between the edge and the spinon is $x$, the total energy of the system is approximately given by:
\begin{align}
E\approx{}[e_{1}(m)-e_{-1}(m)]x+\textrm{const.}\quad{}(m\,\textrm{odd}).
\end{align}
Here $e_{1}(m)$ and $e_{-1}(m)$ are energy per unit length in the even and odd sectors of the YC$4m$ cylinder. Note that $e_{1}>e_{-1}$ when $m$ is odd. Therefore, the spinon is confined to the edge by a linear potential.

Finally, we remark that, as the two topological sectors in YC$4m+2$ cylinders are exactly degenerate, there are no edge spinons in generic conditions.

\section{Discussion}\label{Discussion}

We close our paper by placing our work in the context of previous theoretical efforts. As we have shown in Section \ref{qdm2z2}, the hard core constraints of dimers naturally endow the kagome QDM a $Z_2$ gauge theory structure, by which any kagome QDM Hamiltonian can be formulated as a $Z_2$ lattice gauge theory Hamiltonian. Earlier workers \cite{Nikolic, Huh} used an alternative $Z_2$ gauge-theoretic description of the kagome QDM. In that framework, the dimer occupation number on each kagome link is identical to the $Z_2$ electric flux on that link. The dimer hard-core constraints are approximately enforced by Gauss's law and the energetics. The effective $Z_2$ gauge theory is therefore a $Z_2$ gauge theory on kagome with charge $Q=-1$ on every site.

Despite the superficial differences, the two frameworks are quite similar. In the previously used framework, visons are defined on both triangular and hexagonal plaquettes of kagome. A vison is associated with quantum tunneling  of dimer coverings in a plaquette. However, dimer coverings on a triangular plaquette cannot tunnel as they are eigenstates of the local interactions, and thus it is natural to treat the visons on triangular plaquettes as high energy degrees of freedom and integrate them out. The resulting theory should be the same as ours. Our approach thus provides a more economical representation of the QDM on kagome.

The similarity between the two frameworks becomes more evident in the dual formulation. The dual Landau-Ginzburg order parameter of \textcite{Huh} has four independent real components $\phi_i$. The only quadratic term allowed by the $GL(2,\mathbb{Z}_3)$ symmetry of the dual theory is $\sum_i\phi^2_i$, which is related to the fact that the four sublattices cannot be coupled at the quadratic level as we discussed in Section~\ref{duality_transform}.

After submission of this paper we learned of a similar work by Ju and Balents,\cite{Ju} who mapped a quantum dimer model on kagome onto a $Z_2$ gauge theory on a dice lattice. They computed the dimer modulations in cylinders with various boundary conditions and obtained results in agreement with ours.

\begin{acknowledgments}
The authors would like to thank Leon Balents, Zhihao Hao, Hongchen Jiang, Masaki Oshikawa, Andreas L\"{a}uchli, Matthias Punk, and Steven White for illuminating discussions and for sharing unpublished results. YW acknowledges the hospitality of the Kavli Institute for Theoretical Physics, where a part of the current work was carried out. The research was supported by the U.S. Department of Energy, Office of Basic Energy Sciences, Division of Materials Sciences and Engineering under Award DE-FG02-08ER46544.
\end{acknowledgments}

\bibliography{kagome}

\begin{thebibliography}{31}%
\makeatletter
\providecommand \@ifxundefined [1]{%
 \@ifx{#1\undefined}
}%
\providecommand \@ifnum [1]{%
 \ifnum #1\expandafter \@firstoftwo
 \else \expandafter \@secondoftwo
 \fi
}%
\providecommand \@ifx [1]{%
 \ifx #1\expandafter \@firstoftwo
 \else \expandafter \@secondoftwo
 \fi
}%
\providecommand \natexlab [1]{#1}%
\providecommand \enquote  [1]{``#1''}%
\providecommand \bibnamefont  [1]{#1}%
\providecommand \bibfnamefont [1]{#1}%
\providecommand \citenamefont [1]{#1}%
\providecommand \href@noop [0]{\@secondoftwo}%
\providecommand \href [0]{\begingroup \@sanitize@url \@href}%
\providecommand \@href[1]{\@@startlink{#1}\@@href}%
\providecommand \@@href[1]{\endgroup#1\@@endlink}%
\providecommand \@sanitize@url [0]{\catcode `\\12\catcode `\$12\catcode
  `\&12\catcode `\#12\catcode `\^12\catcode `\_12\catcode `\%12\relax}%
\providecommand \@@startlink[1]{}%
\providecommand \@@endlink[0]{}%
\providecommand \url  [0]{\begingroup\@sanitize@url \@url }%
\providecommand \@url [1]{\endgroup\@href {#1}{\urlprefix }}%
\providecommand \urlprefix  [0]{URL }%
\providecommand \Eprint [0]{\href }%
\providecommand \doibase [0]{http://dx.doi.org/}%
\providecommand \selectlanguage [0]{\@gobble}%
\providecommand \bibinfo  [0]{\@secondoftwo}%
\providecommand \bibfield  [0]{\@secondoftwo}%
\providecommand \translation [1]{[#1]}%
\providecommand \BibitemOpen [0]{}%
\providecommand \bibitemStop [0]{}%
\providecommand \bibitemNoStop [0]{.\EOS\space}%
\providecommand \EOS [0]{\spacefactor3000\relax}%
\providecommand \BibitemShut  [1]{\csname bibitem#1\endcsname}%
\let\auto@bib@innerbib\@empty
\bibitem [{\citenamefont {Balents}(2010)}]{Balents}%
  \BibitemOpen
  \bibfield  {author} {\bibinfo {author} {\bibfnamefont {L.}~\bibnamefont
  {Balents}},\ }\href {\doibase 10.1038/nature08917} {\bibfield  {journal}
  {\bibinfo  {journal} {Nature}\ }\textbf {\bibinfo {volume} {464}},\ \bibinfo
  {pages} {199} (\bibinfo {year} {2010})}\BibitemShut {NoStop}%
\bibitem [{\citenamefont {Kitaev}(2006)}]{Kitaev}%
  \BibitemOpen
  \bibfield  {author} {\bibinfo {author} {\bibfnamefont {A.}~\bibnamefont
  {Kitaev}},\ }\href {\doibase 10.1016/j.aop.2005.10.005} {\bibfield  {journal}
  {\bibinfo  {journal} {Ann. Phys. (N. Y.)}\ }\textbf {\bibinfo {volume}
  {321}},\ \bibinfo {pages} {2} (\bibinfo {year} {2006})}\BibitemShut {NoStop}%
\bibitem [{\citenamefont {Ramirez}(1994)}]{Ramirez}%
  \BibitemOpen
  \bibfield  {author} {\bibinfo {author} {\bibfnamefont {A.~P.}\ \bibnamefont
  {Ramirez}},\ }\href {\doibase 10.1146/annurev.ms.24.080194.002321} {\bibfield
   {journal} {\bibinfo  {journal} {Annu. Rev. Mater. Sci.}\ }\textbf {\bibinfo
  {volume} {24}},\ \bibinfo {pages} {453} (\bibinfo {year} {1994})}\BibitemShut
  {NoStop}%
\bibitem [{\citenamefont {Han}\ \emph {et~al.}(2012)\citenamefont {Han},
  \citenamefont {Helton}, \citenamefont {Chu}, \citenamefont {Nocera},
  \citenamefont {Rodriguez-Rivera}, \citenamefont {Broholm},\ and\
  \citenamefont {Lee}}]{Han}%
  \BibitemOpen
  \bibfield  {author} {\bibinfo {author} {\bibfnamefont {T.-H.}\ \bibnamefont
  {Han}}, \bibinfo {author} {\bibfnamefont {J.~S.}\ \bibnamefont {Helton}},
  \bibinfo {author} {\bibfnamefont {S.}~\bibnamefont {Chu}}, \bibinfo {author}
  {\bibfnamefont {D.~G.}\ \bibnamefont {Nocera}}, \bibinfo {author}
  {\bibfnamefont {J.~A.}\ \bibnamefont {Rodriguez-Rivera}}, \bibinfo {author}
  {\bibfnamefont {C.}~\bibnamefont {Broholm}}, \ and\ \bibinfo {author}
  {\bibfnamefont {Y.~S.}\ \bibnamefont {Lee}},\ }\href {\doibase
  10.1038/nature11659} {\bibfield  {journal} {\bibinfo  {journal} {Nature}\
  }\textbf {\bibinfo {volume} {492}},\ \bibinfo {pages} {406} (\bibinfo {year}
  {2012})}\BibitemShut {NoStop}%
\bibitem [{\citenamefont {Sachdev}(1992)}]{Sachdev}%
  \BibitemOpen
  \bibfield  {author} {\bibinfo {author} {\bibfnamefont {S.}~\bibnamefont
  {Sachdev}},\ }\href {\doibase 10.1103/PhysRevB.45.12377} {\bibfield
  {journal} {\bibinfo  {journal} {Phys. Rev. B}\ }\textbf {\bibinfo {volume}
  {45}},\ \bibinfo {pages} {12377} (\bibinfo {year} {1992})}\BibitemShut
  {NoStop}%
\bibitem [{\citenamefont {Hastings}(2000)}]{Hastings}%
  \BibitemOpen
  \bibfield  {author} {\bibinfo {author} {\bibfnamefont {M.~B.}\ \bibnamefont
  {Hastings}},\ }\href {\doibase 10.1103/PhysRevB.63.014413} {\bibfield
  {journal} {\bibinfo  {journal} {Phys. Rev. B}\ }\textbf {\bibinfo {volume}
  {63}},\ \bibinfo {pages} {014413} (\bibinfo {year} {2000})}\BibitemShut
  {NoStop}%
\bibitem [{\citenamefont {Singh}\ and\ \citenamefont {Huse}(2007)}]{Singh}%
  \BibitemOpen
  \bibfield  {author} {\bibinfo {author} {\bibfnamefont {R.~R.~P.}\
  \bibnamefont {Singh}}\ and\ \bibinfo {author} {\bibfnamefont {D.~A.}\
  \bibnamefont {Huse}},\ }\href {\doibase 10.1103/PhysRevB.76.180407}
  {\bibfield  {journal} {\bibinfo  {journal} {Phys. Rev. B}\ }\textbf {\bibinfo
  {volume} {76}},\ \bibinfo {pages} {180407} (\bibinfo {year}
  {2007})}\BibitemShut {NoStop}%
\bibitem [{\citenamefont {Ran}\ \emph {et~al.}(2007)\citenamefont {Ran},
  \citenamefont {Hermele}, \citenamefont {Lee},\ and\ \citenamefont
  {Wen}}]{Ran}%
  \BibitemOpen
  \bibfield  {author} {\bibinfo {author} {\bibfnamefont {Y.}~\bibnamefont
  {Ran}}, \bibinfo {author} {\bibfnamefont {M.}~\bibnamefont {Hermele}},
  \bibinfo {author} {\bibfnamefont {P.~A.}\ \bibnamefont {Lee}}, \ and\
  \bibinfo {author} {\bibfnamefont {X.-G.}\ \bibnamefont {Wen}},\ }\href
  {\doibase 10.1103/PhysRevLett.98.117205} {\bibfield  {journal} {\bibinfo
  {journal} {Phys. Rev. Lett.}\ }\textbf {\bibinfo {volume} {98}},\ \bibinfo
  {pages} {117205} (\bibinfo {year} {2007})}\BibitemShut {NoStop}%
\bibitem [{\citenamefont {Evenbly}\ and\ \citenamefont
  {Vidal}(2010)}]{Evenbly}%
  \BibitemOpen
  \bibfield  {author} {\bibinfo {author} {\bibfnamefont {G.}~\bibnamefont
  {Evenbly}}\ and\ \bibinfo {author} {\bibfnamefont {G.}~\bibnamefont
  {Vidal}},\ }\href {\doibase 10.1103/PhysRevLett.104.187203} {\bibfield
  {journal} {\bibinfo  {journal} {Phys. Rev. Lett.}\ }\textbf {\bibinfo
  {volume} {104}},\ \bibinfo {pages} {187203} (\bibinfo {year}
  {2010})}\BibitemShut {NoStop}%
\bibitem [{\citenamefont {Yan}\ \emph {et~al.}(2011)\citenamefont {Yan},
  \citenamefont {Huse},\ and\ \citenamefont {White}}]{Yan}%
  \BibitemOpen
  \bibfield  {author} {\bibinfo {author} {\bibfnamefont {S.}~\bibnamefont
  {Yan}}, \bibinfo {author} {\bibfnamefont {D.~A.}\ \bibnamefont {Huse}}, \
  and\ \bibinfo {author} {\bibfnamefont {S.~R.}\ \bibnamefont {White}},\ }\href
  {\doibase 10.1126/science.1201080} {\bibfield  {journal} {\bibinfo  {journal}
  {Science}\ }\textbf {\bibinfo {volume} {332}},\ \bibinfo {pages} {1173}
  (\bibinfo {year} {2011})}\BibitemShut {NoStop}%
\bibitem [{\citenamefont {Depenbrock}\ \emph {et~al.}(2012)\citenamefont
  {Depenbrock}, \citenamefont {McCulloch},\ and\ \citenamefont
  {{Schollw\"{o}ck}}}]{Depenbrock}%
  \BibitemOpen
  \bibfield  {author} {\bibinfo {author} {\bibfnamefont {S.}~\bibnamefont
  {Depenbrock}}, \bibinfo {author} {\bibfnamefont {I.~P.}\ \bibnamefont
  {McCulloch}}, \ and\ \bibinfo {author} {\bibfnamefont {U.}~\bibnamefont
  {{Schollw\"{o}ck}}},\ }\href {\doibase 10.1103/PhysRevLett.109.067201}
  {\bibfield  {journal} {\bibinfo  {journal} {Phys. Rev. Lett.}\ }\textbf
  {\bibinfo {volume} {109}},\ \bibinfo {pages} {067201} (\bibinfo {year}
  {2012})}\BibitemShut {NoStop}%
\bibitem [{\citenamefont {Messio}\ \emph {et~al.}(2012)\citenamefont {Messio},
  \citenamefont {Bernu},\ and\ \citenamefont {Lhuillier}}]{Messio}%
  \BibitemOpen
  \bibfield  {author} {\bibinfo {author} {\bibfnamefont {L.}~\bibnamefont
  {Messio}}, \bibinfo {author} {\bibfnamefont {B.}~\bibnamefont {Bernu}}, \
  and\ \bibinfo {author} {\bibfnamefont {C.}~\bibnamefont {Lhuillier}},\ }\href
  {\doibase 10.1103/PhysRevLett.108.207204} {\bibfield  {journal} {\bibinfo
  {journal} {Phys. Rev. Lett.}\ }\textbf {\bibinfo {volume} {108}},\ \bibinfo
  {pages} {207204} (\bibinfo {year} {2012})}\BibitemShut {NoStop}%
\bibitem [{\citenamefont {Iqbal}\ \emph {et~al.}()\citenamefont {Iqbal},
  \citenamefont {Becca}, \citenamefont {Sorella},\ and\ \citenamefont
  {Poilblanc}}]{Iqbal}%
  \BibitemOpen
  \bibfield  {author} {\bibinfo {author} {\bibfnamefont {Y.}~\bibnamefont
  {Iqbal}}, \bibinfo {author} {\bibfnamefont {F.}~\bibnamefont {Becca}},
  \bibinfo {author} {\bibfnamefont {S.}~\bibnamefont {Sorella}}, \ and\
  \bibinfo {author} {\bibfnamefont {D.}~\bibnamefont {Poilblanc}},\ }\href@noop
  {} {}\Eprint {http://arxiv.org/abs/1209.1858} {arXiv:1209.1858} \BibitemShut
  {NoStop}%
\bibitem [{\citenamefont {Capponi}\ \emph {et~al.}()\citenamefont {Capponi},
  \citenamefont {Chandra}, \citenamefont {Auerbach},\ and\ \citenamefont
  {Weinstein}}]{Capponi}%
  \BibitemOpen
  \bibfield  {author} {\bibinfo {author} {\bibfnamefont {S.}~\bibnamefont
  {Capponi}}, \bibinfo {author} {\bibfnamefont {V.~R.}\ \bibnamefont
  {Chandra}}, \bibinfo {author} {\bibfnamefont {A.}~\bibnamefont {Auerbach}}, \
  and\ \bibinfo {author} {\bibfnamefont {M.}~\bibnamefont {Weinstein}},\
  }\href@noop {} {}\Eprint {http://arxiv.org/abs/1210.5519} {arXiv:1210.5519}
  \BibitemShut {NoStop}%
\bibitem [{\citenamefont {Jiang}\ \emph
  {et~al.}(2012{\natexlab{a}})\citenamefont {Jiang}, \citenamefont {Wang},\
  and\ \citenamefont {Balents}}]{Jiang}%
  \BibitemOpen
  \bibfield  {author} {\bibinfo {author} {\bibfnamefont {H.-C.}\ \bibnamefont
  {Jiang}}, \bibinfo {author} {\bibfnamefont {Z.}~\bibnamefont {Wang}}, \ and\
  \bibinfo {author} {\bibfnamefont {L.}~\bibnamefont {Balents}},\ }\href
  {\doibase doi:10.1038/nphys2465} {\bibfield  {journal} {\bibinfo  {journal}
  {Nature Physics}\ }\textbf {\bibinfo {volume} {8}},\ \bibinfo {pages} {902}
  (\bibinfo {year} {2012}{\natexlab{a}})}\BibitemShut {NoStop}%
\bibitem [{\citenamefont {White}()}]{White}%
  \BibitemOpen
  \bibfield  {author} {\bibinfo {author} {\bibfnamefont {S.~R.}\ \bibnamefont
  {White}},\ }\href@noop {} {}\bibinfo {howpublished} {unpublished}\BibitemShut
  {NoStop}%
\bibitem [{\citenamefont {Jiang}\ and\ \citenamefont {Balents}()}]{Jiang2}%
  \BibitemOpen
  \bibfield  {author} {\bibinfo {author} {\bibfnamefont {H.-C.}\ \bibnamefont
  {Jiang}}\ and\ \bibinfo {author} {\bibfnamefont {L.}~\bibnamefont
  {Balents}},\ }\href@noop {} {}\bibinfo {howpublished}
  {unpublished}\BibitemShut {NoStop}%
\bibitem [{\citenamefont {Moessner}\ \emph {et~al.}(2001)\citenamefont
  {Moessner}, \citenamefont {Sondhi},\ and\ \citenamefont
  {Fradkin}}]{Moessner2}%
  \BibitemOpen
  \bibfield  {author} {\bibinfo {author} {\bibfnamefont {R.}~\bibnamefont
  {Moessner}}, \bibinfo {author} {\bibfnamefont {S.~L.}\ \bibnamefont
  {Sondhi}}, \ and\ \bibinfo {author} {\bibfnamefont {E.}~\bibnamefont
  {Fradkin}},\ }\href {\doibase 10.1103/PhysRevB.65.024504} {\bibfield
  {journal} {\bibinfo  {journal} {Phys. Rev. B}\ }\textbf {\bibinfo {volume}
  {65}},\ \bibinfo {pages} {024504} (\bibinfo {year} {2001})}\BibitemShut
  {NoStop}%
\bibitem [{\citenamefont {Moessner}\ and\ \citenamefont
  {Sondhi}(2001)}]{Moessner1}%
  \BibitemOpen
  \bibfield  {author} {\bibinfo {author} {\bibfnamefont {R.}~\bibnamefont
  {Moessner}}\ and\ \bibinfo {author} {\bibfnamefont {S.~L.}\ \bibnamefont
  {Sondhi}},\ }\href {\doibase 10.1103/PhysRevLett.96.110405} {\bibfield
  {journal} {\bibinfo  {journal} {Phys. Rev. Lett.}\ }\textbf {\bibinfo
  {volume} {86}},\ \bibinfo {pages} {1881} (\bibinfo {year}
  {2001})}\BibitemShut {NoStop}%
\bibitem [{\citenamefont {Jiang}\ \emph
  {et~al.}(2012{\natexlab{b}})\citenamefont {Jiang}, \citenamefont {Yao},\ and\
  \citenamefont {Balents}}]{J1J2}%
  \BibitemOpen
  \bibfield  {author} {\bibinfo {author} {\bibfnamefont {H.-C.}\ \bibnamefont
  {Jiang}}, \bibinfo {author} {\bibfnamefont {H.}~\bibnamefont {Yao}}, \ and\
  \bibinfo {author} {\bibfnamefont {L.}~\bibnamefont {Balents}},\ }\href
  {\doibase 10.1103/PhysRevB.86.024424} {\bibfield  {journal} {\bibinfo
  {journal} {Phys. Rev. B}\ }\textbf {\bibinfo {volume} {86}},\ \bibinfo
  {pages} {024424} (\bibinfo {year} {2012}{\natexlab{b}})}\BibitemShut
  {NoStop}%
\bibitem [{\citenamefont {Misguich}\ \emph {et~al.}(2002)\citenamefont
  {Misguich}, \citenamefont {Serban},\ and\ \citenamefont
  {Pasquier}}]{Misguich}%
  \BibitemOpen
  \bibfield  {author} {\bibinfo {author} {\bibfnamefont {G.}~\bibnamefont
  {Misguich}}, \bibinfo {author} {\bibfnamefont {D.}~\bibnamefont {Serban}}, \
  and\ \bibinfo {author} {\bibfnamefont {V.}~\bibnamefont {Pasquier}},\ }\href
  {\doibase 10.1103/PhysRevLett.89.137202} {\bibfield  {journal} {\bibinfo
  {journal} {Phys. Rev. Lett.}\ }\textbf {\bibinfo {volume} {89}},\ \bibinfo
  {pages} {137202} (\bibinfo {year} {2002})}\BibitemShut {NoStop}%
\bibitem [{\citenamefont {Nikolic}\ and\ \citenamefont
  {Senthil}(2003)}]{Nikolic}%
  \BibitemOpen
  \bibfield  {author} {\bibinfo {author} {\bibfnamefont {P.}~\bibnamefont
  {Nikolic}}\ and\ \bibinfo {author} {\bibfnamefont {T.}~\bibnamefont
  {Senthil}},\ }\href {\doibase 10.1103/PhysRevB.68.214415} {\bibfield
  {journal} {\bibinfo  {journal} {Phys. Rev. B}\ }\textbf {\bibinfo {volume}
  {68}},\ \bibinfo {pages} {214415} (\bibinfo {year} {2003})}\BibitemShut
  {NoStop}%
\bibitem [{\citenamefont {Huh}\ \emph {et~al.}(2011)\citenamefont {Huh},
  \citenamefont {Punk},\ and\ \citenamefont {Sachdev}}]{Huh}%
  \BibitemOpen
  \bibfield  {author} {\bibinfo {author} {\bibfnamefont {Y.}~\bibnamefont
  {Huh}}, \bibinfo {author} {\bibfnamefont {M.}~\bibnamefont {Punk}}, \ and\
  \bibinfo {author} {\bibfnamefont {S.}~\bibnamefont {Sachdev}},\ }\href
  {\doibase 10.1103/PhysRevB.84.094419} {\bibfield  {journal} {\bibinfo
  {journal} {Phys. Rev. B}\ }\textbf {\bibinfo {volume} {84}},\ \bibinfo
  {pages} {094419} (\bibinfo {year} {2011})}\BibitemShut {NoStop}%
\bibitem [{\citenamefont {Zeng}\ and\ \citenamefont {Elser}(1995)}]{Zeng}%
  \BibitemOpen
  \bibfield  {author} {\bibinfo {author} {\bibfnamefont {C.}~\bibnamefont
  {Zeng}}\ and\ \bibinfo {author} {\bibfnamefont {V.}~\bibnamefont {Elser}},\
  }\href {\doibase 10.1103/PhysRevB.51.8318} {\bibfield  {journal} {\bibinfo
  {journal} {Phys. Rev. B}\ }\textbf {\bibinfo {volume} {51}},\ \bibinfo
  {pages} {8318} (\bibinfo {year} {1995})}\BibitemShut {NoStop}%
\bibitem [{\citenamefont {Elser}\ and\ \citenamefont {Zeng}(1993)}]{Elser}%
  \BibitemOpen
  \bibfield  {author} {\bibinfo {author} {\bibfnamefont {V.}~\bibnamefont
  {Elser}}\ and\ \bibinfo {author} {\bibfnamefont {C.}~\bibnamefont {Zeng}},\
  }\href {\doibase 10.1103/PhysRevB.48.13647} {\bibfield  {journal} {\bibinfo
  {journal} {Phys. Rev. B}\ }\textbf {\bibinfo {volume} {48}},\ \bibinfo
  {pages} {13647} (\bibinfo {year} {1993})}\BibitemShut {NoStop}%
\bibitem [{\citenamefont {Rokhsar}\ and\ \citenamefont
  {Kivelson}(1988)}]{Rokhsar}%
  \BibitemOpen
  \bibfield  {author} {\bibinfo {author} {\bibfnamefont {D.~S.}\ \bibnamefont
  {Rokhsar}}\ and\ \bibinfo {author} {\bibfnamefont {S.~A.}\ \bibnamefont
  {Kivelson}},\ }\href {\doibase 10.1103/PhysRevLett.61.2376} {\bibfield
  {journal} {\bibinfo  {journal} {Phys. Rev. Lett.}\ }\textbf {\bibinfo
  {volume} {61}},\ \bibinfo {pages} {2376} (\bibinfo {year}
  {1988})}\BibitemShut {NoStop}%
\bibitem [{\citenamefont {Kogut}(1979)}]{Kogut}%
  \BibitemOpen
  \bibfield  {author} {\bibinfo {author} {\bibfnamefont {J.}~\bibnamefont
  {Kogut}},\ }\href {\doibase 10.1103/RevModPhys.51.659} {\bibfield  {journal}
  {\bibinfo  {journal} {Rev. Mod. Phys.}\ }\textbf {\bibinfo {volume} {51}},\
  \bibinfo {pages} {659} (\bibinfo {year} {1979})}\BibitemShut {NoStop}%
\bibitem [{\citenamefont {Savit}(1980)}]{Savit}%
  \BibitemOpen
  \bibfield  {author} {\bibinfo {author} {\bibfnamefont {R.}~\bibnamefont
  {Savit}},\ }\href {\doibase 10.1103/RevModPhys.52.453} {\bibfield  {journal}
  {\bibinfo  {journal} {Rev. Mod. Phys.}\ }\textbf {\bibinfo {volume} {52}},\
  \bibinfo {pages} {453} (\bibinfo {year} {1980})}\BibitemShut {NoStop}%
\bibitem [{\citenamefont {Lieb}\ \emph {et~al.}(1961)\citenamefont {Lieb},
  \citenamefont {Schultz},\ and\ \citenamefont {Mattis}}]{Lieb}%
  \BibitemOpen
  \bibfield  {author} {\bibinfo {author} {\bibfnamefont {E.}~\bibnamefont
  {Lieb}}, \bibinfo {author} {\bibfnamefont {T.}~\bibnamefont {Schultz}}, \
  and\ \bibinfo {author} {\bibfnamefont {D.}~\bibnamefont {Mattis}},\ }\href
  {\doibase 10.1016/0003-4916(61)90115-4} {\bibfield  {journal} {\bibinfo
  {journal} {Ann. Phys. (N. Y.)}\ }\textbf {\bibinfo {volume} {16}},\ \bibinfo
  {pages} {407} (\bibinfo {year} {1961})}\BibitemShut {NoStop}%
\bibitem [{\citenamefont {{L\"{a}uchli}}()}]{Lauchli}%
  \BibitemOpen
  \bibfield  {author} {\bibinfo {author} {\bibfnamefont {A.}~\bibnamefont
  {{L\"{a}uchli}}},\ }\href@noop {} {}\bibinfo {howpublished}
  {unpublished}\BibitemShut {NoStop}%
\bibitem [{\citenamefont {Ju}\ and\ \citenamefont {Balents}()}]{Ju}%
  \BibitemOpen
  \bibfield  {author} {\bibinfo {author} {\bibfnamefont {H.}~\bibnamefont
  {Ju}}\ and\ \bibinfo {author} {\bibfnamefont {L.}~\bibnamefont {Balents}},\
  }\href@noop {} {}\Eprint {http://arxiv.org/abs/1302.2636} {arXiv:1302.2636}
  \BibitemShut {NoStop}%
\end{thebibliography}%

\end{document}